\documentclass[preprint,a4paper,showkeys,superscriptaddress]{revtex4}
\usepackage{graphicx}

\begin{document}

\newcommand{\be}{\begin{equation}}
\newcommand{\ee}{\end{equation}}
\newcommand{\bq}{\begin{eqnarray}}
\newcommand{\eq}{\end{eqnarray}}
\newcommand{\bsq}{\begin{subequations}}
\newcommand{\esq}{\end{subequations}}
\newcommand{\bc}{\begin{center}}
\newcommand{\ec}{\end{center}}
\newcommand{\al}{\alpha}

\title{Velocity-Dependent Models for Non-Abelian/Entangled String Networks}

\author{A. Avgoustidis}
\email[Electronic address: ]{tasos@ecm.ub.es}
\affiliation{Departament d'Estructura i Costituents de la Mat\`eria,\\ 
Universitat de Barcelona, Diagonal 647, 08028 Barcelona, Spain}
\affiliation{Department of Applied Mathematics and Theoretical Physics,
Centre for Mathematical Sciences,\\ University of Cambridge,
Wilberforce Road, Cambridge CB3 0WA, United Kingdom}
\author{E.P.S. Shellard}
\email[Electronic address: ]{E.P.S.Shellard@damtp.cam.ac.uk}
\affiliation{Department of Applied Mathematics and Theoretical Physics,
Centre for Mathematical Sciences,\\ University of Cambridge,
Wilberforce Road, Cambridge CB3 0WA, United Kingdom}

\begin{abstract}
We develop velocity-dependent models describing the evolution of string
networks that involve several types of interacting strings, each
with a different tension.   These incorporate the formation of Y-type 
junctions with links stretching between colliding strings, while 
always ensuring energy conservation.  These models can be used to 
describe network evolution for non-abelian strings as well as cosmic 
superstrings.  The application to $Z_{N}$ strings in which interactions 
are topologically constrained, demonstrates that a scaling regime is 
generally reached which involves a hierarchy of string densities with 
the lightest most abundant.  We also study hybrid networks of cosmic 
superstrings, where energetic considerations are more important in 
determining interaction outcomes.   We again find that networks 
tend towards scaling, with the three lightest network components 
being dominant and having comparable number densities, while the 
heavier string states are suppressed.  A more quantitative analysis 
depends on the precise calculation of the string interaction matrix 
using the underlying string or field theory.  Nevertheless, these 
results provide further evidence that the presence of junctions in 
a string network does not obstruct scaling.
\end{abstract}


\preprint{}
\maketitle

\section{\label{intro}Introduction}

 Much of the interest in cosmic strings was lost with the realisation that  
 they cannot be the dominant seeds for structure formation in the Universe   
 \cite{Battye,Battye1}.  However, their appearance in many cosmological   
 situations forces one to consider them as subdominant contributors.   
 Recent theoretical work in brane inflation \cite{DvalTye,Quevedo,KKLMMT}   
 and SUSY GUTs \cite{Jeannerot}, as well as the potential for observing 
 strings through gravitational lensing \cite{Sazhin,Sazhin1,Schild}
 and the CMB \cite{LoWright}, have contributed to a significant revival 
 of interest in the subject (for reviews see for example  
 Refs.~\cite{Kibblerev,PolchIntro,DavKib,Majumdar_tut}).  More recent 
 work~\cite{BatGarMos,BHindKU} suggests that CMB data allow a 10-15\% 
 contribution from cosmic strings, arguably favouring a scale-invariant 
 spectrum plus strings over a tilted spectrum without strings. 

 Particularly interesting is that cosmic strings can be produced in 
 the final stages of brane inflation \cite{DvalTye,DvalShafSolg,BMNQRZ, 
 Garc-Bell,JoStoTye1,KKLMMT}, as D-branes and fundamental strings  
 stretching over cosmological distances \cite{SarTye,DvalVil,PolchStab}.   
 These objects, often referred  to as `cosmic superstrings', can have  
 different phenomenology than ordinary field  theory strings.  This opens  
 up the possibility that cosmic string observations could yield information  
 about physics at the string scale, and provides new ways of constraining   
 various brane inflation models \cite{Babichev,cycloops,BvdBDD,Melkumova}.   
 Cosmic superstrings have small tensions (in the range $10^{-12} < G\mu  
 < 10^{-7}$ \cite{SarTye,JoStoTye2,PolchIntro}) and reconnect with  
 probabilities that can be significantly less than unity  
 \cite{JoStoTye2,PolchProb}.  This leads to a reduced intercommutation  
 rate, resulting in an enhancement of the predicted string number density  
 today \cite{JoStoTye2,DvalVil,EDVOS}.  Cosmic superstring networks can 
 consist of more than one type of string \cite{DvalTye,PolchStab}, which 
 can zip together to produce trilinear vertices with links (or better,  
 in the terminology of Ref.~\cite{McGraw1}, zippers) stretching between  
 them.  In this respect  they are similar to non-abelian strings  
 \cite{book,SperPen,McGraw,McGraw1}.  In contrast to abelian string  
 networks, which have been shown to always reach a scaling solution  
 \cite{Kibble,Bennett,AlbTur,AllShel,vosk}, non-abelian strings could  
 in principle be frustrated \cite{SperPen}.   

 The evolution of non-abelian string networks was further studied in    
 \cite{McGraw,McGraw1} where some evidence for scaling behaviour was found.   
 There have also been more recent attempts to model non-abelian strings  
 both analytically and numerically \cite{nonint,MTVOS,CopSaf,CopKibSteer1,
 CopKibSteer2,Saffin,HindSaf}, favouring scaling (see also  
 Ref.~\cite{wallscaling} where scaling was found for domain wall networks  
 with junctions).  In this paper we present a class of velocity-dependent  
 models for non-abelian string evolution, developed under significantly  
 different assumptions than those of Ref.~\cite{MTVOS}.  In particular,  
 rather than associating a different energy density to each type of string,  
 while keeping a single correlation length and average velocity for  
 all types, we associate a different correlation length and velocity   
 to each string type and assume a Brownian network structure, in which   
 correlation lengths also quantify string energy densities.  We consider  
 two basic categories of interactions between strings of different types,  
 namely the coalescence along their own length (zipping) that  
 produces a `zipper', and the creation of a new segment (a `bridge')  
 of different string type as the strings pass through one another.
 The relative importance of these inter-related dynamical mechanisms 
 will depend on the details of the field theory model under study, 
 and more general interactions can be described as relative combinations
 of these two limiting cases, as we shall comment later.  Our network 
 evolution models are constructed in a manner close to the traditional 
 (Kibble) approach, where the new terms corresponding to the production 
 of bridges/zippers, are introduced through energetic considerations 
 of string interactions within a certain volume element.  We express 
 our models in terms of the 3D energy density of strings, thus ensuring 
 the use of well established techniques in calculating the energy loss 
 of the long string network.  We can then directly compare our results 
 to both the usual (3D) VOS model \cite{vosk} and numerical simulations  
 \cite{BenBouch,AllShel}. 
     
 Central to our models is ensuring that energy is conserved by balancing 
 the energies corresponding to the produced and lost string lengths
 and/or the kinetic energies of the interacting and produced string 
 segments.  In this way we avoid the possibility of finding `spurious
 scaling' by artificially throwing away energy from regions of 
 intercommutation.  Realistic collisions can of course be inelastic, in 
 that some of the energy can be released in the form of radiation.  
 This provides an additional energy loss mechanism, which facilitates 
 the reaching of a scaling regime.  In the absence of a quantitative 
 understanding of the importance of such an energy loss mechanism, we 
 take a conservative approach and assume energy balance at the 
 interaction regions.  We find that even in this case, where this 
 potentially important energy loss mechanism is effectively switched 
 off, multi-tension networks with junctions can still reach scaling.   
 
 The structure of the paper is as follows: In section \ref{evolution} 
 we briefly review some basic facts about cosmic string evolution both   
 in the abelian and non-abelian case. In section \ref{NAVOS} we move on   
 to construct a class of velocity-dependent models for non-abelian string  
 evolution.  Application to $Z_N$ strings is presented in section  
 \ref{Z_Nstrings}, whereas in section \ref{super} we consider the case  
 of cosmic superstrings.  Our conclusions are summarised in section  
 \ref{conc}.

\section{\label{evolution}Cosmic String Evolution}

  \subsection{Abelian Strings}
   
   Most well-studied examples of cosmic strings are those arising in  
   theories with a broken $U(1)$ symmetry, like the Nielsen-Olesen vortex 
   lines of the abelian-Higgs model. Understanding the interaction of   
   strings when they intersect is crucial for modelling their cosmic   
   evolution.  For such $U(1)$ strings, crossing has two topologically   
   allowed outcomes: either the strings pass through one another, or they   
   intercommute (reconnect) by exchanging partners.  Which situation  
   is actually realised is a dynamical question.  This problem has been   
   extensively studied numerically in field theory~\cite{Shell_int, Matzner},  
   where it was found that the strings reconnect with probability of   
   order unity\footnote{See Ref.~\cite{Achuc} for a recent study of 
   possible exceptions to this rule in high-velocity regimes.}.  Some 
   recent analytic results have also been presented in Ref.~\cite{HanHash}.  
   Having quantitative understanding of string-string interactions is 
   necessary for modelling the cosmological evolution of strings.  Below 
   we review some basic results.     
 
   \subsubsection{\label{basics}Basics} 
    Numerical simulations of string formation in field theory suggest  
    that strings formed after cosmological phase transitions have, to 
    a good approximation, the shape of random walks. This allows one 
    to describe a string network by a   
    characteristic length $L$, which determines both the typical radius    
    of curvature of strings and the average interstring distance. There   
    is typically one string segment of length $L$ in each volume $L^3$    
    so that the energy density of the network can be defined as   
    \be\label{rho}
     \rho=\frac{\mu L}{L^3}=\frac{\mu}{L^2} \,.
    \ee
      
    As the network evolves, strings collide with each other or curl back   
    on themselves creating small loops, which oscillate and radiatively   
    decay. Via these interactions enough energy is lost from the network    
    to ensure that the strings do not dominate the energy density of the   
    universe\footnote{Energy loss through production of particles could 
    also play a significant role~\cite{VinHindSak,MooShelMart}.} . An 
    approximate energy loss rate equation can be written   
    \cite{Kibble}  
    \be\label{rholoss}
     \dot\rho\approx -2\frac{\dot a}{a}\rho - \frac{\rho}{L}\, ,
    \ee  
    where the first term is due to Hubble expansion ($a=a(t)$ being the  
    scalefactor of the universe) and the second term models the energy   
    lost through string interactions and the formation of loops. Such   
    interacting networks are known to evolve towards a so-called `scaling'    
    regime in which the characteristic length $L$ stays constant relative    
    to the the horizon $d_H\sim t$ \cite{Kibble}. This can be seen by    
    setting $L=\gamma(t) t$ and substituting (\ref{rho}) into    
    (\ref{rholoss}) to obtain
    \be\label{gdot}
     \frac{\dot\gamma}{\gamma}=\frac{1}{2t}\left(2(\beta-1)+\frac{1}{\gamma}
     \right) \,.
    \ee

    The expansion exponent $\beta$ is related to the scalefactor $a(t)$   
    by $a(t) \propto t^\beta$ and is equal to $1/2$ and $2/3$ in the   
    radiation and matter eras respectively. Equation (\ref{gdot}) has   
    indeed a scaling solution
    \be\label{slnosm}
     \gamma=[2(1-\beta)]^{-1}
    \ee
    demonstrating that the characteristic length asymptotically reaches 
    a constant value with respect to the horizon $L\sim t$. 
    If one starts with a high density
    of strings, intercommuting will produce loops reducing the energy of   
    the network, whereas if the initial density is low then there will not   
    be enough intercommuting and $\gamma$ will decrease. Given enough time,    
    the two competing effects of stretching and fragmentation will always   
    reach a steady-state and the scaling regime will be approached.
     
    The above discussion captures the key physical processes involved  
    in string evolution, but for a more quantitative study one needs models  
    of higher sophistication. String evolution has been extensively studied  
    numerically through Nambu-Goto simulations~\cite{AlbTur,BenBouch,AllShel}, 
    verifying the above picture, but a number of analytic models have 
    also been developed. These include   
    a velocity-dependent one-scale (VOS) model \cite{vos0,vos,vosk}, a   
    `kink-counting' model \cite{AllCald,Austin}, a functional approach   
    \cite{Embacher}, a `three-scale' model \cite{AusCopKib} and a `wiggly'  
    model \cite{thesis}. In the following we will consider the VOS model,   
    as it is the simplest of these and has been shown to be in very good   
    agreement with numerical simulations \cite{vostests}.    

   \subsubsection{\label{VOS}The VOS model}
    The velocity-dependent one-scale (VOS) model is a simple analytic model,  
    depending on one free parameter only, which agrees quantitatively with  
    Nambu-Goto string evolution simulations.  From the theoretical 
    point of view it is  
    well-motivated and can be obtained directly from the Nambu-Goto action,  
    by performing a statistical averaging procedure on the string equations   
    of motion and energy momentum tensor along the network \cite{vos0}.
    In particular, from the energy momentum tensor one can obtain an   
    evolution equation for the string energy density $\rho$, whereas   
    the Nambu equations of motion yield a `macroscopic' equation for the   
    evolution of the typical rms velocity $v$ of string segments.      

    The resulting equations are:
    \be\label{rhodtvos}
     \dot\rho = -2\frac{\dot a}{a}(1+v^2)\rho-\frac{\tilde c v\rho}{L}
    \ee
    \be\label{vdtvos}
     \dot v = (1-v^2)\left(\frac{k}{R}-2\frac{\dot a}{a}v\right)
    \ee   
 
    The second term in equation (\ref{rhodtvos}) is a phenomenological term   
    which takes into account the energy losses through creation of loops.   
    This depends on the loop production parameter $\tilde c$, related to
    the integral of an appropriate loop production function over all
    relevant loop sizes \cite{book}.  In equation (\ref{vdtvos}), $R$ is 
    the average radius of curvature of strings in the network and $k$  
    the so-called momentum parameter (see later discussion) introduced  
    in Ref.~\cite{vos0}.  For a Brownian network, and within the VOS  
    assumptions, the average radius of curvature $R$ can be taken equal  
    to the correlation length $L\equiv\gamma t$.  
                                                                   
    Using (\ref{rho}) and setting $L=\gamma(t)\,t$ as before, we
    obtain the following equation instead of (\ref{rhodtvos}):
    \be\label{gdtvos}
     \frac{\dot\gamma}{\gamma}=\frac{1}{2t}\left(2\beta(1+v^2)-
     2+\frac{\tilde c v}{\gamma}\right)\, .
    \ee
                                                                          
    This is of the same form as (\ref{gdot}) but has an extra correction   
    term $\beta v^2$ accounting for redshifting of velocities due to   
    cosmological expansion. It also includes the parameter $\tilde c$,   
    the value of which can be extracted from numerical simulations and   
    it is of order unity \cite{book,vos}.  Possibly reduced intercommuting 
    probabilities relevant to cosmic superstrings, or non-intercommuting 
    for high velocities in type II strings~\cite{Achuc}, correspond to 
    a smaller effective value of $\tilde c$.   
                                                                          
    The system (\ref{vdtvos}-\ref{gdtvos}) has the scaling solution
    \be\label{solnvos}
      \gamma^2=\frac{k(k+\tilde c)}{4\beta(1-\beta)}  \;\;\;\;\;\;\;\;\;\;
       v^2=\frac{k(1-\beta)}{\beta(k+\tilde c)}
    \ee
    in terms of the expansion exponent $\beta$, the loop production
    parameter $\tilde c$, and the momentum parameter $k$ which is 
    a measure of the smoothness of the strings.  An accurate ansatz 
    for the momentum parameter has been proposed in \cite{vosk}
    \be\label{kans3d}
     k = k(v) = \frac{2\sqrt{2}}{\pi} \left( \frac{1-8 v^6}{1+8 v^6}
     \right) \,,
    \ee
    which incorporates the `Virial' condition $v^2\le 1/2$, observed in 
    Nambu-Goto simulations \cite{book}.  

    With this ansatz the VOS model depends on one parameter only, $\tilde c$, 
    which, as mentioned above, can readily be extracted by comparison to  
    numerical simulations.  Remarkably, by adjusting this parameter once 
    only, one can closely model high resolution Nambu-Goto simulations 
    throughout cosmic history.

  \subsection{\label{non_abn}Non-Abelian Strings}
 
   Consider a situation where a gauge symmetry group $G$ is spontaneously
   broken to a subgroup $H$. This process can give rise to topological
   defects classified by the homotopy groups of the vacuum manifold $G/H$.
   In particular, string defects can occur if the vacuum manifold is not
   simply connected that is $\pi_1(G/H) \ne I$, where $\pi_1$ is the
   fundamental group and $I$ the identity. In the example of $U(1)$
   strings the symmetry breaking is $U(1)\rightarrow I$ so $\pi_1(G/H)=Z$,
   the group of integers, corresponding to the integer winding number
   of $U(1)$ strings. For a connected and simply connected gauge group,
   the fundamental theorem implies $\pi_1(G/H) \sim \pi_0(H)$, where
   $\pi_0(H)$ is the set of disconnected components of $H$. Thus, for
   a simply connected $G$, strings can occur if $H$ is disconnected.

   The flux $f$ of a string is given by a path-ordered exponential of the
   gauge field \cite{ALM-RP}
   \be\label{flux}
    f = P \exp \left( i \oint_\gamma A_{\mu} \, dx^{\mu} \right)\,,
   \ee
   where $\gamma$ is a closed, oriented path, which encloses the string
   without passing from the string core and without enclosing any other
   strings. The requirement that the Higgs field which drives the symmetry
   breaking is invariant under parallel transport along such a closed
   path forces the flux to be an element of the unbroken group $H$. If
   $H$ is non-abelian then the above definition is not gauge invariant:
   under a gauge transformation by $h \in H$ the flux $f$ changes to
   $hfh^{-1}$. However, two paths which start and end at the same point
   $p$ and which can be continuously deformed to each other necessarily
   have the same flux.

\begin{figure}[h]
     \includegraphics[height=2.5in,width=4.5in]{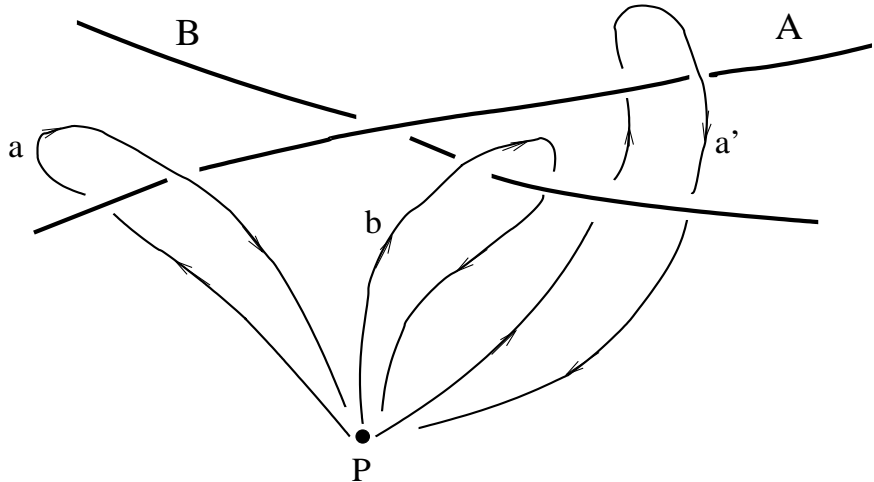}
     \caption{\label{stringsAB} The paths $\alpha$ and $\alpha^{\prime}$
              on the two sides of the strings are not homotopically
              equivalent and as a result the corresponding fluxes are
              related by conjugation with $\beta$, that is
              $\alpha^{\prime}=\beta \alpha \beta^{-1}$.}
    \end{figure}

   The non-abelian structure of $H$ gives rise to a certain degree of
   ambiguity in defining the flux of a particular string. Consider a
   string A which is entangled to another string B as in
   Fig.~\ref{stringsAB}.
   One can calculate the flux $\alpha$ on one side of $A$ by
   considering the path $a$, and the flux on the other side
   $\alpha^{\prime}$ by considering path $a^{\prime}$.  Since
   strings A and B are entangled, path $a$ cannot be deformed to
   $a^{\prime}$ without encountering string B.  Therefore, they
   correspond to different elements of the fundamental group and the
   fluxes $\alpha$ and $\alpha^{\prime}$ are generally different.
   In fact they are  related by conjugation with the flux $\beta$
   (of string B, corresponding to path $b$) that is
   \be\label{aprimea}
    \alpha^{\prime}=\beta \alpha \beta^{-1} \, ,
   \ee
   which gives $\alpha^{\prime}=\alpha$ in the abelian case.  This
   is not merely a mathematical curiosity, but a physical effect
   that can be interpreted in terms of a long-range interaction
   between non-commuting strings, the so called `holonomy interaction'
   \cite{Bucher,Wilc-Wu,ALM-RP}.  Thus, the flux of a particular string
   is not associated to a unique element but can be described by
   several distinct conjugate elements of $H$.  It follows that
   strings with fluxes in the same conjugacy class must have the
   same tension.

   Non-commutativity of $H$ changes the nature of string interactions,
   thus giving rise to an interesting property of non-abelian strings:
   the ability to form \emph{entangled} networks. Indeed, two strings
   carrying non-commuting fluxes can neither pass through each other
   nor reconnect, as both possibilities would violate flux conservation
   \cite{Toulouse,Mermin}.
   The collision of two such strings leads to the formation of a new
   string segment stretching between the two, which carries flux given
   by the commutator of the initial fluxes. A related process is
   that of \emph{branching}, that is the splitting of a string with
   flux $c=ab$ to two strings with fluxes $a$ and $b$. This is not
   however special to having a non-abelian unbroken group $H$; it can
   also occur for strings arising through the process $U(1)\rightarrow
   Z_N$, with $N\ge 3$ \cite{McGraw,AEVV,VachVil}.  Also, for type I 
   $U(1)$ strings, colliding segments can zip for small enough crossing 
   angle and velocity of approach \cite{Bettencourt}.  

   The evolution of entangled and/or branched networks is not as well
   understood as that of ordinary $U(1)$ string networks. A certain
   class of non-abelian strings have been studied numerically in
   Ref.~\cite{SperPen}, where the possibility of a `frustrated'
   network (one that does not exhibit the scaling property and
   eventually dominates the energy density of the universe) was
   identified. Simulations of $S_3$ strings were presented in
   Refs.~\cite{McGraw,McGraw1}, where strong evidence for scaling
   behaviour was found. Interest in the subject has recently revived
   in the context of `cosmic superstring' networks, which also have
   the branching property. Recent attempts to model such networks both
   analytically and numerically can be found in
   Refs.~\cite{nonint,MTVOS,CopSaf,Saffin,HindSaf}. In particular
   Ref.~\cite{nonint} suggests a way to generalise the VOS model to
   take into account the formation of links (or `bridges') in an
   entangled/branched network.  Ref.~\cite{MTVOS} proposes a further
   generalisation in order to model networks composed by different
   types of strings, each with different tension.  Finally,
   Refs.~\cite{CopSaf,Saffin,HindSaf} take a field theory approach
   to cosmic superstring models.

   The aim of this paper is to develop an analytic model for the evolution
   of multi-string, entangled/branched networks in the spirit of
   Refs.~\cite{nonint,MTVOS}.  Our work closely follows the traditional
   Kibble approach to modelling string-string interactions in terms
   of energetic considerations in a correlation length volume.  We start
   developing our models in the next section.

\section{\label{NAVOS}Non-Abelian VOS models}  

 We wish to study a network of cosmic strings consisting of several   
 different types of string, possibly with different tensions and   
 non-commuting fluxes. The two mechanisms governing the evolution  
 of the network are the cosmological expansion (with its associated  
 velocity redshift effects) and the string-string interactions. For  
 U(1) strings both effects are well-described by the VOS model, but   
 to allow for the new possibilities arising in a multi-string network   
 (section \ref{non_abn}) this model needs to be accordingly modified.   
 As in Ref.~\cite{MTVOS} we aim to have a set of VOS equations for   
 each type of string, with extra terms added in order to describe the   
 interactions between different string types which transfer energy   
 from one type to another.  

\begin{figure}[t] 
    \begin{center} 
    \includegraphics[height=2.5cm,width=11cm]{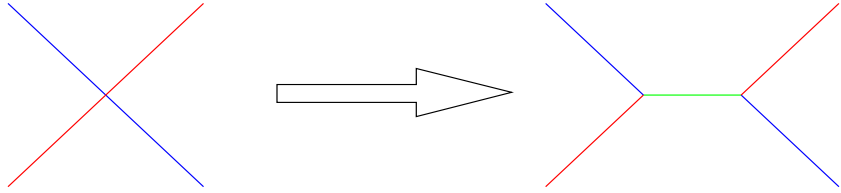}
    \end{center}
    \caption{\label{bridge} The bridge-type interaction: the  
             colliding strings form a new segment, the `bridge',  
             as they pass through each other.} 
  \end{figure}

\begin{figure}[!h] 
    \begin{center}
    \includegraphics[height=3cm,width=12cm]{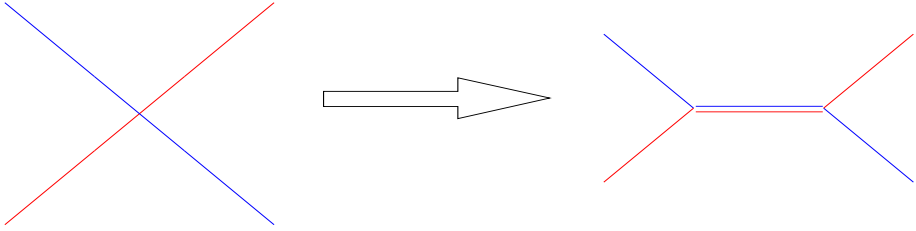}
    \end{center}
    \caption{\label{zipper} The zipper-type interaction: strings 
        zip along part of their own length to form a `zipper'.}
  \end{figure}

 String interactions are modelled as follows: when two strings of the  
 \emph{same type} collide there are two topologically allowed outcomes;  
 either they intercommute or pass through one another. This will be  
 described by the usual VOS parameter $\tilde c$ (see Eq.~(\ref{rhodtvos})),  
 parametrising the energy lost through self-intersection of strings and  
 the subsequent production of loops of the same type. However, if the  
 collision involves \emph{different types} of strings (with non-commuting   
 fluxes) then both the above possibilities violate flux conservation.   
 The collision instead leads to the production of a new string segment,  
 joining the two strings~\footnote{Here we neglect the more complex  
 mechanism by which loops consisting of several types of string are  
 lost to the network.}.  As already mentioned we choose to describe 
 the formation of the new segment via two inter-related limiting 
 mechanisms: 
 \medskip\\
 \noindent {\it (i)} The strings pass through one another and in doing so  
           they become linked by a new string segment of a third type  
           (Fig.~\ref{bridge}).  This is the so-called `bridge'  
           configuration in the terminology of Ref.~\cite{McGraw1}. 
           In this asymptotic limit the energy to make the new link  
           is acquired from the kinetic energy of the original pair.        
 \medskip\\
 \noindent {\it (ii)} The strings coalesce along part of their own length,  
           in the so-called `zipper' configuration of Fig.~\ref{zipper}. 
           In this limit there must be significant energy gain from the 
           realignment and coalescence of the two original strings, which 
           lose physical length to create the new segment.   
  \smallskip\\ 
 A general model will involve some combination of these two 
 limiting mechanisms.  Their relative importance is both a dynamical 
 and kinematic question depending on the string tensions, velocities 
 and angles between the colliding strings, as well as topological 
 considerations.  Model-dependent averages, such as those considered 
 in Refs.~\cite{CopKibSteer1,CopKibSteer2}, may be used to determine  
 their relative quantitative importance.  We will first consider the  
 impact of each of these possibilities separately in sections \ref{bridges}  
 and \ref{zippers}.  Then we will present the general case with both  
 mechanisms incorporated to describe more general string interactions.       
 
 \subsection{\label{example}Toy Models}   
   
  Consider a simple example where we begin with a network of Brownian   
  strings of types $1$ and $2$, with tensions $\mu_1$, $\mu_2$ and  
  correlation lengths $L_1$, $L_2$ respectively. We can think of this     
  system as two separate networks, each characterised by its own   
  correlation length, which are allowed to interact with each other.  
  Interactions between strings of the same type are as in the $U(1)$  
  case, but when strings of type $1$ and $2$ collide they produce   
  a new segment of type $3$ with tension $\mu_3$.  We assume that the 
  energy density of segments of type 3 can be described by a correlation  
  length $L_3$, as for a Brownian network.  We can then define the  
  energy density $\rho_3=\mu_3/L_3^2$ and write an evolution equation 
  \be\label{rho3dt}   
   \dot\rho_3 = -2\frac{\dot a}{a}(1+v_3^2)\rho_3-\frac{\tilde c_3 v_3  
   \rho_3}{L_3} + \dot\rho_{1,2\rightarrow 3} \,.  
  \ee    
  The first two terms are as in Eq.~(\ref{rhodtvos}) and describe the  
  effects of cosmological expansion and loop production through   
  self-intersection of strings of type $3$. The third term models   
  the production of new segments of type $3$ string from interactions   
  between strings of types $1$ and $2$.  

  To write down a formula for $\dot\rho_{1,2\rightarrow 3}$ we consider   
  a correlation-length-sized segment of type $1$ string interacting   
  with the type $2$ network and take, without loss of generality, $L_1<L_2$.  
  On average there is one type $2$ segment of length $L_2$ in each  
  correlation volume $L_2^3$, so that, if $\bar v$ is the velocity of  
  the type $1$ segment relative to the (type $2$) network, then the  
  probability that the segment meets a string of type $2$ in time  
  $\delta t$ is:  
  \be\label{prob_meet}  
   \frac{\bar v \delta t}{L_2} \frac{L_1}{L_2} \,.     
  \ee
  Such a collision results in the production of a new type $3$ segment. In  
  analogy to the case of loop production, we will assume that the length   
  distribution of the produced links is peaked at a value $\ell(t)$.   
  Integrating over this distribution function introduces an efficiency    
  parameter $\tilde d$, the analogue of parameter $\tilde c$ for the case  
  of loop production.  Considering that on average there are $\frac{\rho_1} 
  {\mu_1 L_1}L_2^3=L_2^3/L_1^3$ strings of type $1$ in each cube of volume  
  $L_2^3$ we can write  
  \be\label{rhodot12t3}  
   \dot\rho_{1,2\rightarrow 3}=\tilde d \frac{\bar v}{L_2}   
   \frac{L_1}{L_2} \frac{\mu_3 \ell(t)}{L_2^3} \frac{L_2^3}{L_1^3} 
   =\frac{\tilde d \bar v \mu_3 \ell(t)}{L_1^2 L_2^2} \,.  
  \ee
  Note that this expression is symmetric in $L_1, L_2$ as expected.   
     
  With equation (\ref{rhodot12t3}) modelling the production of links  
  due to interactions of strings of different type, and with the  
  assumption that the produced links form a Brownian network, we can  
  go on to construct simple VOS models of non-abelian string evolution.  
  We will first consider toy models of networks consisting of three 
  types of string only, which interact in a simple manner, and in 
  the next sections we will progressively build more complex models.
  Before presenting the models we briefly discuss the important issue
  of energy conservation.  

  \subsubsection{\label{en_cons}Energy Conservation Issues}
  When considering macroscopic string evolution equations, it is  
  important to make sure that energy is conserved by the interactions.    
  In the 3D VOS model, the loop production term damps energy from the  
  long string network, but this energy is simply transferred to the   
  network of cosmic loops.  However, for processes where two colliding   
  long string segments produce a (long) segment of third type, it is   
  important that the energy  differences balance exactly, so that no 
  net energy is produced or lost.  If this condition is not met, one 
  is in danger of unphysically throwing energy away, which may lead 
  to a `spurious' scaling regime.  Of course, string collisions may 
  well be inelastic, and it seems plausible that some energy can, 
  indeed, be lost through particle production~\cite{MooShelMart,VinHindSak}
  or gravitational radiation.  However, it is at present not clear 
  what proportion of the 
  relevant energies can indeed be damped away from the network through, 
  say, massive radiation.  In the case of Abelian-Higgs networks, there
  is evidence that massive radiation is subdominant compared to loop 
  production and gravitational radiation decay mechanisms~\cite{MooShelMart}.
 
  In the absence of a quantitative understanding of particle production 
  in the present context, it seems dangerous to assume that the energy 
  difference associated to the zipping, say, process is all damped away 
  through this decay channel.  A more conservative approach would be 
  to impose energy conservation during string intercommutations (in 
  other words to switch off particle production), and see whether 
  the resulting networks can still reach a scaling regime.  That 
  is the approach we will follow in this paper.  As we will see, 
  for interactions of the bridge type, this energy 
  balance can be accomplished by taking into account  
  the slowing down of the original strings due to the tension of the   
  bridge between them, or, in the zipper case, by giving kinetic energy   
  to the produced zipper through terms in the corresponding velocity   
  evolution equations.  We now examine each of these situations 
  separately.  

  \subsubsection{\label{bridges}Model with Bridges}    
  Now consider the case where strings of type $1$ and $2$ produce  
  type $3$ segments through interactions of the `bridge' type   
  (Fig.~\ref{bridge}).  As the strings pass through one another, a new  
  segment of type $3$ is produced, while the initial length of each  
  of the original strings is preserved (Note that by `length' we mean 
  physical, not invariant, length).  The tension of the formed  
  link slows down the original strings, and in doing so its length  
  increases.  If $\ell(t)$ is the average value of the lengths of  
  the links at the end of all interactions which occurred at times  
  around $t$, then the evolution of the energy density of the type  
  $3$ network is given by Eqs.~(\ref{rho3dt}) and (\ref{rhodot12t3}).  
  However, since the length of type $1$ and $2$ strings is not  
  affected by the production of bridges, the corresponding evolution  
  equations for networks $1$ and $2$ will be of the form of  
  Eq.~(\ref{rhodtvos}).  We will therefore have for the energy  
  densities:  
  \be\label{rho_idtbridge}
    \dot\rho_i = -2\frac{\dot a}{a}(1+v_i^2)\rho_i-\frac{\tilde c_i 
    v_i\rho_i}{L_i}, \,\,\,\,\,\, i=1,2
  \ee
  \be\label{rho_3dtbridge} 
    \dot\rho_3 = -2\frac{\dot a}{a}(1+v_3^2)\rho_3-\frac{\tilde c_3 
    v_3\rho_3}{L_3} + \frac{\tilde d \bar v \mu_3 \ell(t)}{L_1^2 L_2^2}\,.  
  \ee  
  
  Equations~(\ref{rho_idtbridge})-(\ref{rho_3dtbridge}) seem to violate  
  energy conservation since for $\dot a=0$ (flat space) and $\tilde c_i=0$ 
  (no loop production), the first two equations become $\dot\rho_i=0$ but 
  $\dot\rho_3$ has a positive term, appearing to create energy from nothing.  
  This energy is actually coming from the kinetic energies of the colliding 
  strings (which have been slowed down due to the tension of the bridge 
  stretching between them) and needs to be taken into account in the  
  velocity equations.  To do this, we express the string density $\rho$ 
  in terms of the proper correlation length $L_0$, taking into account  
  relativistic length contraction of a moving string in a box of size  
  $L_0^3$:  
  \be\label{rhoL0} 
   \rho=\frac{g(\mu L_0)}{L_0^3}=g\frac{\mu}{L_0^2}\, ,   
  \ee   
  where $g\!=\!(1-v^2)^{-1/2}$ is the Lorentz factor associated with  
  the motion of the string.  We can then convert changes in string  
  energy density to string accelerations via the equation 
  \be\label{rho_acc_dt}
   \dot\rho_{\rm acc}=\frac{\partial\rho}{\partial v}
   \frac{{\rm d}v}{{\rm d}t}=\frac{v}{(1-v^2)}\rho \dot v \,.
  \ee      
  We have one such term for each string of type $1$ and $2$ and we
  require that they balance the energy gain due to the production
  of bridges, that is 
  \be\label{bridge_ebal}
   \dot\rho_{1,\rm acc} + \dot\rho_{2,\rm acc} =
   \dot\rho_{1,2\rightarrow3} \,.
  \ee
  Since the strings can have different tensions, the weighting depends  
  on the relative string tensions~\footnote{One could also consider  
  dependence on correlation lengths and/or string velocities.}, so we  
  introduce two weighting factors $w_1(\mu_1,\mu_2)$, $w_2(\mu_1,\mu_2)$  
  such that $w_1(\mu_1,\mu_2)+w_2(\mu_1,\mu_2)=1$ and require
  \be\label{weighted1}
   \dot\rho_{1,\rm acc}=w_1(\mu_1,\mu_2)\dot\rho_{1,2\rightarrow3}
  \ee
  \be\label{weighted2}
   \dot\rho_{2,\rm acc}=w_2(\mu_1,\mu_2)\dot\rho_{1,2\rightarrow3}\,. 
  \ee
  Using Eqs.~(\ref{rho_acc_dt}) and (\ref{rhodot12t3}) we find
  \be\label{v_iterms}
   \dot v_i = \tilde d (1-v_i^2) w_i(\mu_1,\mu_2) \frac{\bar v}{v_i}
   \frac{\mu_3}{\mu_i} \frac{\ell(t)}{L_j^2}, \,\,\,\,\,\, i=1,2 \,\, 
   {\rm and} \,\, i\ne j 
  \ee 
  which are the new terms we seek. The velocity evolution equations
  for our model are therefore
  \be\label{v_idtbridge}  
    \dot v_i = (1-v_i^2)\left(\frac{k_i}{R_i}-2\frac{\dot a}{a}v_i
    -\tilde d w_i(\mu_1,\mu_2) \frac{\bar v}{v_i}\frac{\mu_3}{\mu_i}   
    \frac{\ell(t)}{L_j^2}\right), \,\,\,\,\,\, i=1,2 \,\, {\rm and} 
    \,\, i\ne j 
  \ee 
  \be\label{v_3dtbridge}
    \dot v_3 = (1-v_3^2)\left(\frac{k_3}{R_3}-2\frac{\dot a}{a}v_3
    \right) \,.  
  \ee  
  
\begin{figure}[t]
    \includegraphics[height=2.8in,width=2.9in]{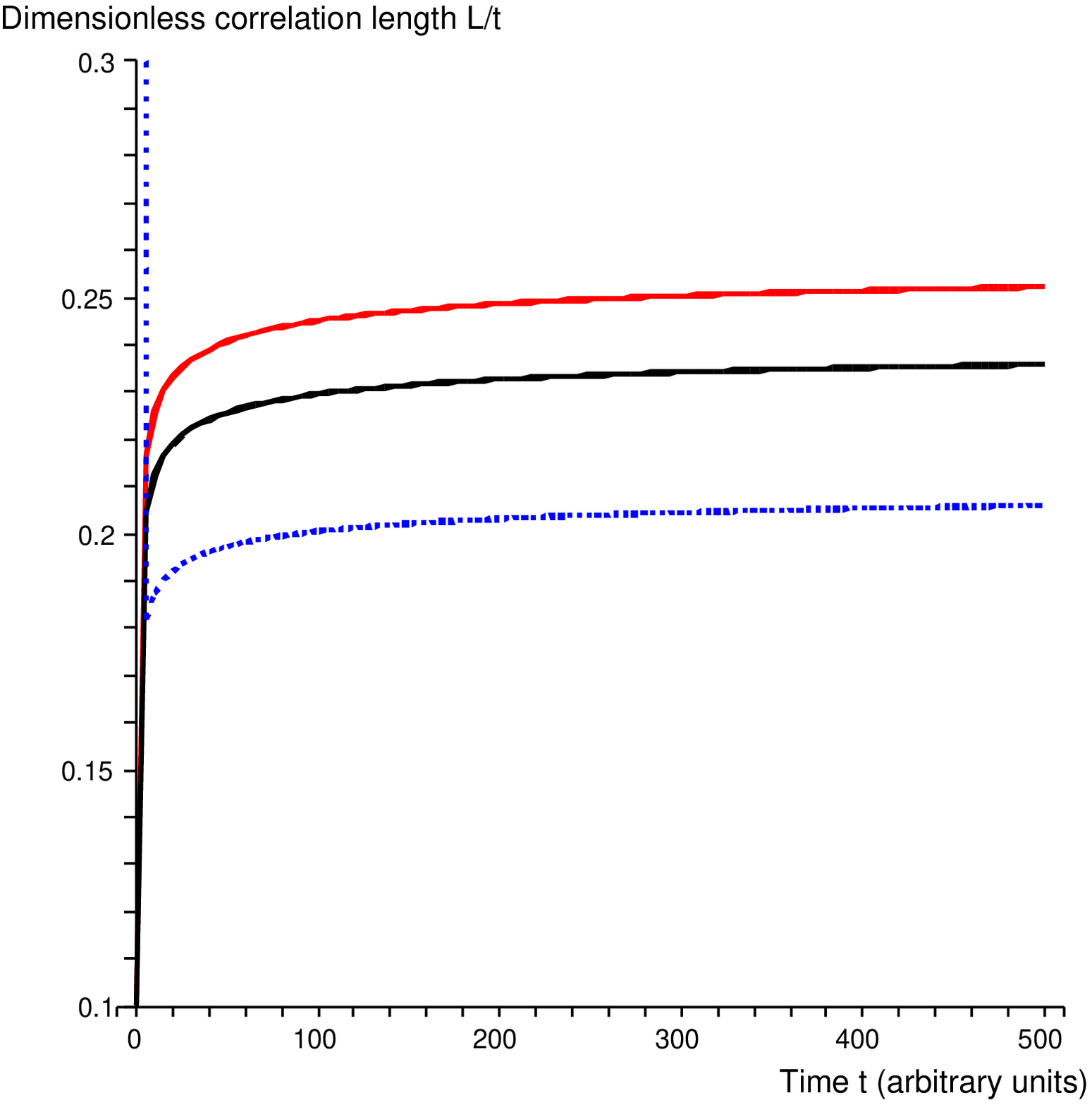} 
    \includegraphics[height=2.8in,width=2.9in]{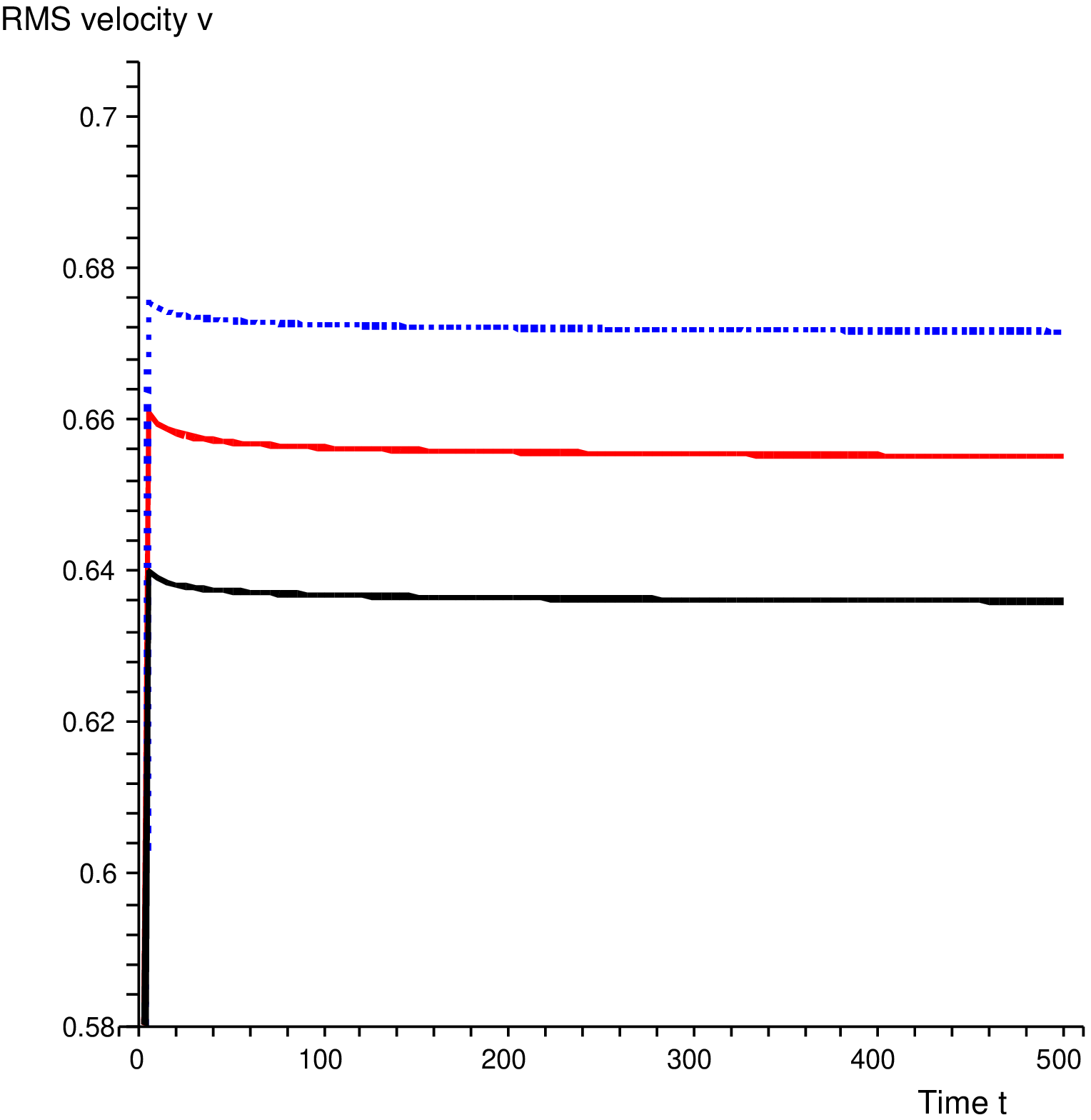}
    \caption{\label{bridge12g} Evolution of $\gamma=L/t$ and 
             $v$ for the three components of the network (bridge 
             model).  Type 1 strings are shown in black, 2 in red  
             and 3 in dotted blue.  Type 3 are more heavily  
             populated as they are continuously produced by  
             interactions.  Strings of type 2 are the less  
             populated, being more massive than type 1,  
             $\mu_2=2\mu_1$.}
   \end{figure}

  This picture is not complete, as we have not specified what happens   
  when a string of type $3$ collides with strings of type $1$ or  
  $2$.  For this toy model we will simply assume that the strings  
  simply pass through one another in such an encounter, but more 
  complicated models of this type will be considered in section  
  \ref{Z_Nstrings}.  Our aim here is to describe how such multi-string 
  models may be constructed and to demonstrate that the presence of  
  different string types in the network, interacting with each other 
  through terms like (\ref{rhodot12t3}), does not necessarily obstruct  
  string scaling.  Indeed, taking all loop production parameters  
  $\tilde c_1$, $\tilde c_2$ and $\tilde c_3$ equal to the canonical 
  value $\tilde c=0.23$ (coming from Nambu-Goto simulation of a single
  string network~\cite{vos}) and solving Eqs.  
  (\ref{rho_idtbridge})-(\ref{rho_3dtbridge}),  
  (\ref{v_idtbridge})-(\ref{v_3dtbridge}) numerically, we find   
  scaling behaviour for all three string types.  This is illustrated 
  in Fig.~\ref{bridge12g} for $\mu_2=2\mu_1$, $\mu_3=\sqrt{\mu_1^2+ 
  \mu_2^2}$.  For the weighting factors we have taken 
  $w_1=\mu_2/(\mu_1+\mu_2)$ and $w_2=\mu_1/(\mu_1+\mu_2)$, since  
  the light string's motion is affected more due to its smaller  
  inertia.  We see that in this model, type $3$ strings (shown
  in a blue dotted line) have smaller $\gamma$ and therefore become
  more abundant than types $1$ and $2$.  This is simply a consequence
  of the fact that we have not allowed the breaking of type $3$ strings
  into types $1$ and $2$: type $3$ strings are continuously produced
  by interactions between type $1$ and $2$ strings, while there is no
  such mechanism to produce string types $1$ or $2$.  Also note the
  important differentiation between string types 1 and 2.  This is 
  because of the velocity dependence (see Eq. (\ref{v_idtbridge})), 
  since type 1 strings with less inertia lose more kinetic energy 
  in an interaction.  Lower velocities imply lower interaction rates 
  for loop production which enhances the overall density of the light 
  string.          
  
  From the point of view of string density evolution, a scaling solution  
  is generically reached, as can be seen directly from the corresponding 
  $\gamma_3=L_3/t$ equation: 
   \be\label{gamma3_scale} 
    \frac{\dot\gamma_3}{\gamma_3} = \frac{1}{2t}
    \left[2\beta(1+v_3^2)-2+\frac{\tilde c_3 v_3}{\gamma_3} 
    - \tilde d \bar v \frac{\gamma_3^2}{\gamma_1^2\gamma_2^2} 
    \frac{\ell(t)}{t} \right] . 
  \ee
  Each term in the right hand side of (\ref{gamma3_scale}) has a different  
  dependence on $\gamma_3$.  Further, the loop production term tends to 
  increase $\gamma_3$, but as this happens this term weakens.  Similarly, 
  the term corresponding to the production of bridges tends to reduce  
  $\gamma_3$ and in doing so it also weakens.  Thus, there is always 
  a value of $\gamma_3$ for which the right hand side of (\ref{gamma3_scale}) 
  is zero (scaling) and the dynamics is such that the system is driven 
  towards that value.  However, if we allow interactions between type  
  3 and type 1 or 2 strings by including bridge production terms in  
  equations (\ref{rho_idtbridge}), then decreasing of $\gamma_1$, $\gamma_2$  
  in the bridge production term of equation (\ref{gamma3_scale}) could in 
  principle lead to a frustrated network.  It is therefore important  
  to consider more complicated situations separately, which we will do
  in section \ref{Z_Nstrings}.      

  \subsubsection{\label{zippers}Model with Zippers}         
  Now consider the case where collisions between strings of type $1$
  and $2$ produce type $3$ segments through interactions of the
  `zipper' type (Fig.~\ref{zipper}). In this case the original   
  strings coalesce along their own length, so there will be 
  extra terms in the corresponding energy density equations,  
  describing the energy loss associated to the length of the colliding  
  string which was converted to type $3$. The energy density evolution  
  equations thus become: 
  \be\label{rho_idtzipper}
     \dot\rho_i = -2\frac{\dot a}{a}(1+v_i^2)\rho_i-\frac{\tilde c_i 
     v_i\rho_i}{L_i}-\frac{\tilde d \bar v \mu_i \ell(t)}{L_1^2
     L_2^2}, \,\,\,\,\,\, i=1,2
   \ee
   \be\label{rho_3dtzipper} 
     \dot\rho_3 = -2\frac{\dot a}{a}(1+v_3^2)\rho_3-\frac{\tilde c_3 
     v_3\rho_3}{L_3} + \frac{\tilde d \bar v \mu_3 \ell(t)}{L_1^2 L_2^2}\,.   
  \ee 
  In general $\mu_1+\mu_2 \ne \mu_3$ so there will be an energy
  density of
  \be\label{deltarho_int}
   \delta\rho_{\rm int} = \tilde d \bar v (\mu_1+\mu_2-\mu_3)
   \frac{\ell(t)}{L_1^2 L_2^2} \delta t 
  \ee
  left from the interaction.  We impose energy conservation by
  demanding that this energy excess is given as kinetic energy to
  the produced zippers, that is we require  
  \be\label{zipper_ebal}
    \dot\rho_{\rm int}=\dot\rho_{3,\rm acc} \, .
  \ee
  This corresponds to the introduction of a new source term in the
  velocity equation of type $3$ strings.  The velocity evolution
  equations for this model are then
  \be\label{v_idt_zipper}
    \dot v_i = (1-v_i^2)\left(\frac{k_i}{R_i}-2\frac{\dot a}{a}v_i
    \right) \,\,\,\,\,\, i=1,2 
  \ee 
  \be\label{v_3dt_zipper}
    \dot v_3 = (1-v_3^2)\left(\frac{k_3}{R_3}-2\frac{\dot a}{a}v_3
    +\tilde d \frac{\bar v}{v_3}\frac{\mu_1+\mu_2-\mu_3}{\mu_3}
    \frac{ L_3^2 \ell(t)}{L_1^2 L_2^2} \right) \,.  
  \ee  
  As explained above (section~\ref{en_cons}), imposing energy 
  conservation at zipping is a conservative approach, which corresponds 
  to switching off the possibility of damping energy through particle 
  production or gravitational radiation.  In this way we avoid the 
  possibility of artificially 
  throwing away too much energy, which could lead to an unphysical 
  scaling regime.  The main purpose of this paper is to demonstrate 
  that loop production alone can still provide an efficient damping 
  mechanism in the presence of links, so that we will assume energy
  conservation throughout the paper.  When using phenomenological 
  models of the type we present here to fit simulations, one can 
  easily relax this assumption by introducing a tunable unbalance 
  through reduction of the coefficient $\tilde d$ in equation 
  (\ref{v_3dt_zipper}). 
  \begin{figure}[!t]
    \includegraphics[height=2.7in,width=2.9in]{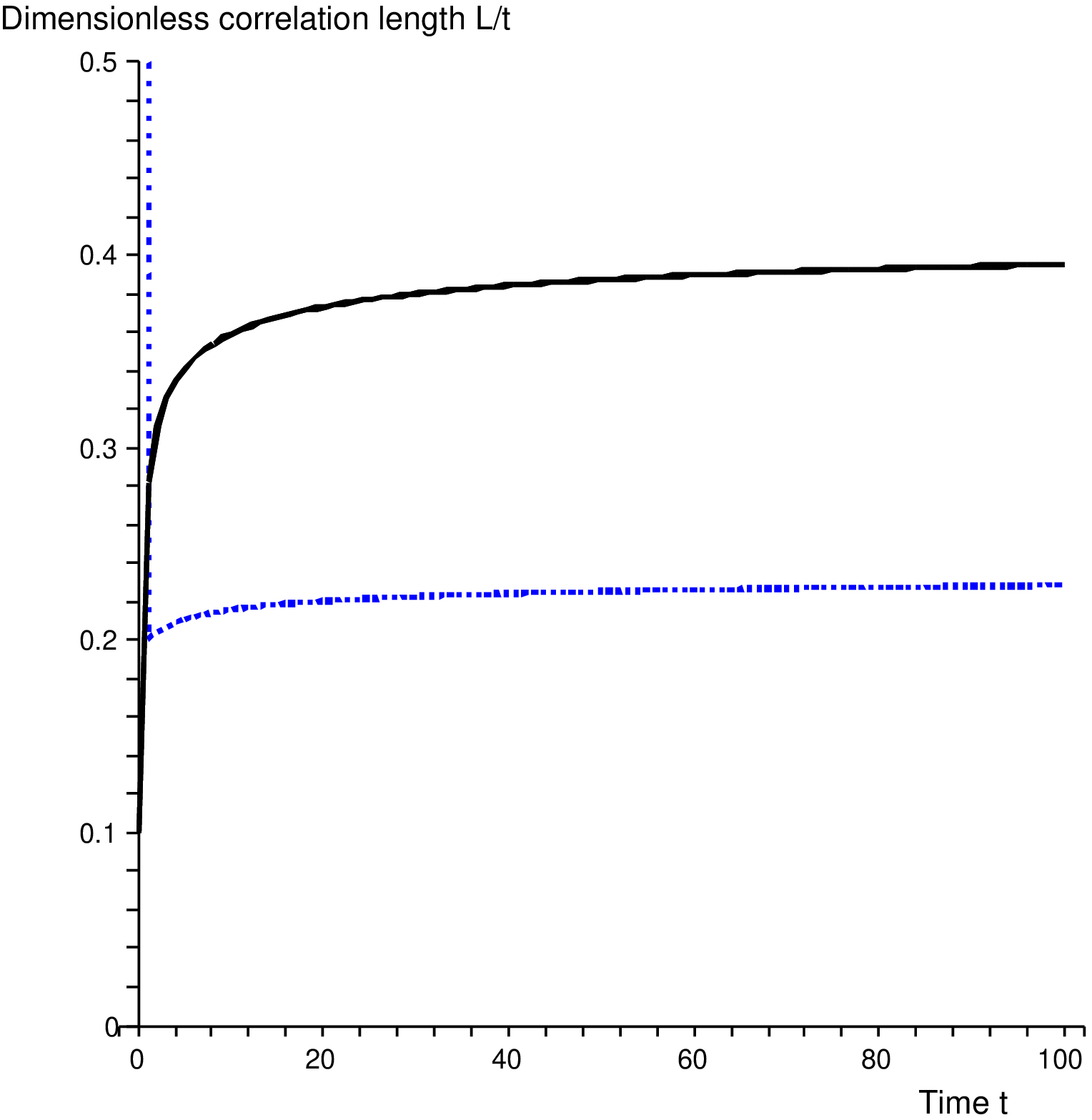}
    \includegraphics[height=2.7in,width=2.9in]{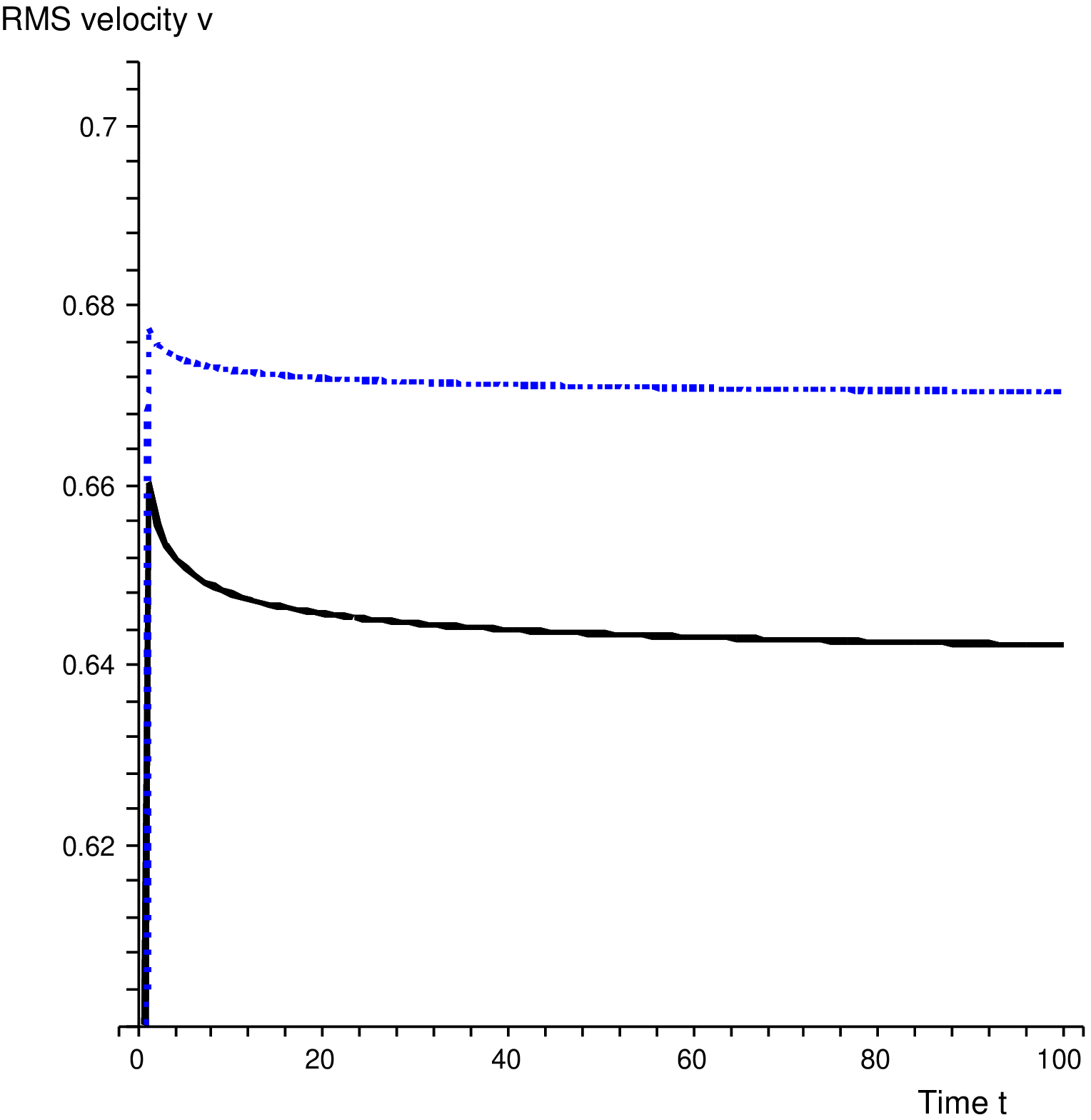}
    \caption{\label{zip12nobreak} Evolution of $\gamma=L/t$ and
             velocity $v$ for the three components of the network
             (zipper model), for the case where the heavy (type 3)
             strings do not interact with the lighter ones.  Type 1
             and 2 strings are shown with a black solid line, and
             type 3 is shown with a blue dotted line.  All three
             components reach scaling with type 3 having smaller
             $\gamma$ (and so higher density) due to its continuous
             production from zipping of type 1 and 2 strings.}
   \end{figure}

\begin{figure}[h]
    \includegraphics[height=2.7in,width=2.9in]{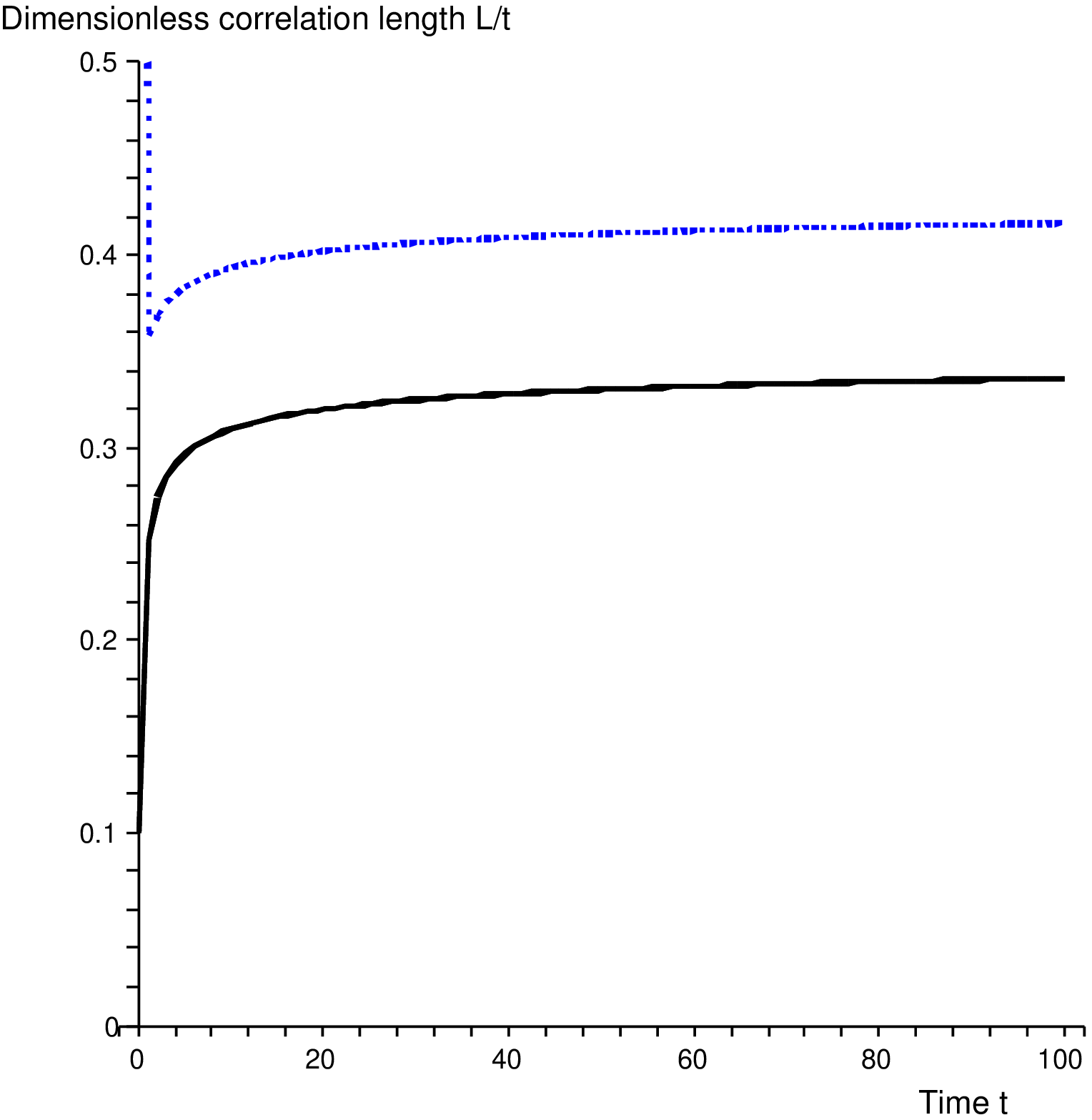}
    \includegraphics[height=2.7in,width=2.9in]{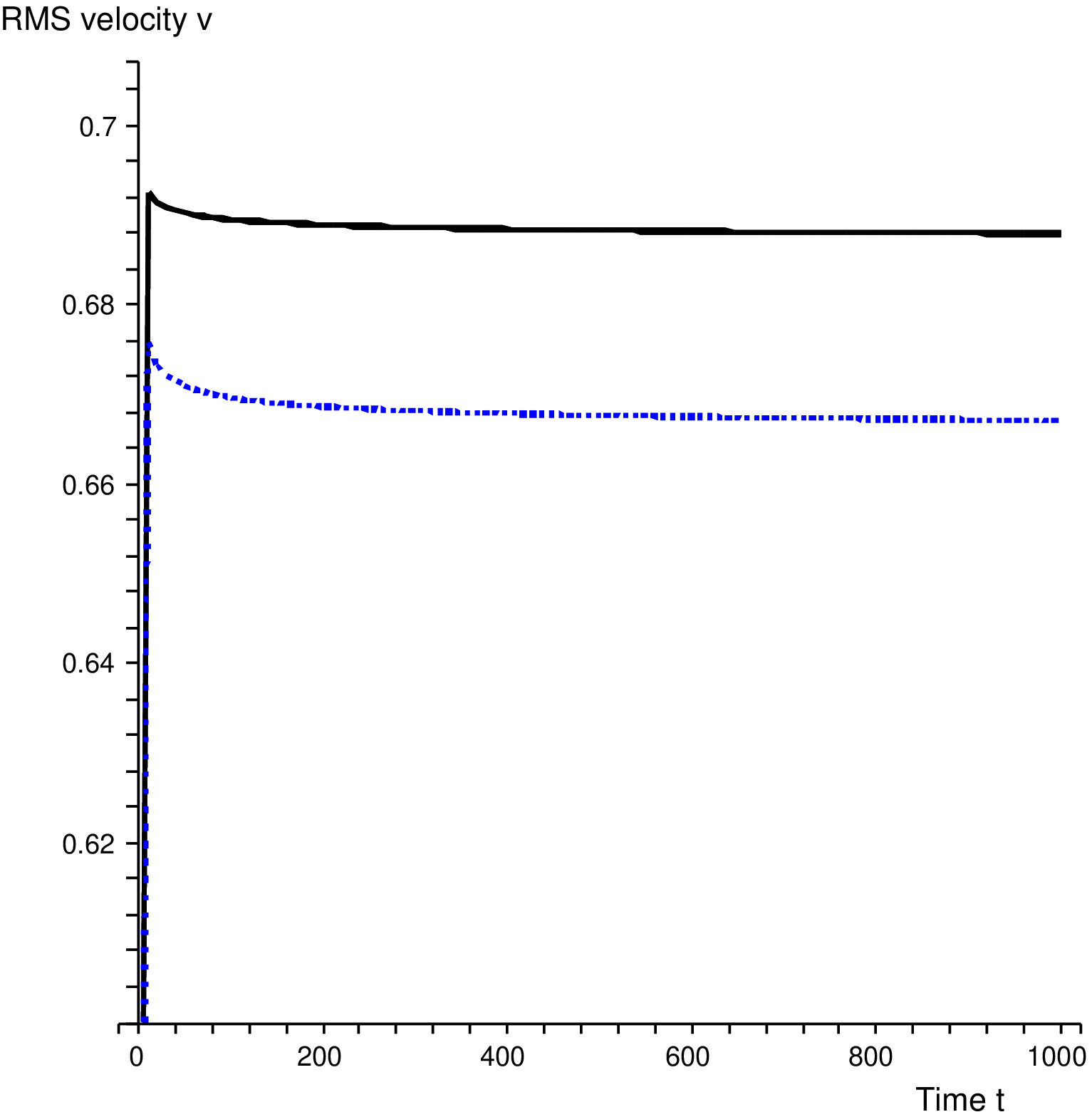}
    \caption{\label{zip12andbreak} An example of the evolution of  
             $\gamma$ and $v$ for a network (zipper model) where the  
             heavy type 3 strings can interact with the lighter ones  
             and unzip.  The heavy type 3 strings with  
             $\mu_3=\sqrt{\mu_1+\mu_2}$ are shown with a dotted  
             blue line and the lighter strings with a black solid 
             one.  For simplicity we have taken $\mu_1=\mu_2$.  All  
             string types reach scaling but now type 3 strings can  
             have smaller density due to their unzipping.}
   \end{figure}

  Solving Eqs.~(\ref{rho_idtzipper})-(\ref{rho_3dtzipper}),  
  (\ref{v_idt_zipper})-(\ref{v_3dt_zipper}) numerically with $\tilde  
  c_1=\tilde c_2=\tilde c_3=0.23$ we find scaling for all string types    
  (Fig.~\ref{zip12nobreak}), where, as before, the heaviest type $3$  
  strings are more highly populated.  Again, this is because we have not  
  included the possibility of `unzipping' of type $3$ strings through  
  their collisions with type $1$ or $2$: string length is continuously  
  lost from type $1$ and $2$ strings into forming type $3$ zippers,  
  but the type $3$ network only loses length through loop production.   
  Allowing for the possibility that strings of type $3$ collide with  
  type $1$ (resp. type $2$) producing type $2$ (resp. type $1$)  
  segments one also finds scaling solutions were the most massive 
  type 3 strings are less heavily populated (Fig.~\ref{zip12andbreak}).   
  We see therefore in this toy model that all string interactions  
  have to be taken carefully into account, in order to obtain the  
  correct scaling.  We will construct more complicated models of 
  this type in section \ref{super}.      

  
  \subsection{\label{general}General Case} 
  Arbitrarily complex models of multi-string networks can be built 
  in a similar way by replicating the VOS equations   
  (\ref{rhodtvos}),(\ref{vdtvos}) and adding extra terms of the type  
  discussed above. The result is a tower of ODE's of the form 
  \be\label{rho_idtgen} 
    \dot\rho_i = -2\frac{\dot a}{a}(1+v_i^2)\rho_i-\frac{\tilde c_i 
    v_i\rho_i}{L_i} - \sum_{a,k}^{\rm zippers} \frac{\tilde d_{ia}^k   
    \bar v_{ia} \mu_i \ell_{ia}^k(t)}{L_a^2 L_i^2} + \sum_{b,\,a\le b}   
    \frac{\tilde d_{ab}^i \bar v_{ab} \mu_i   
    \ell_{ab}^i(t)}{L_a^2 L_b^2}        
  \ee  
  \bq
    \dot v_i = (1-v_i^2)\left[\frac{k_i}{R_i}-2\frac{\dot a}{a}v_i 
    -\sum_{a,k}^{\rm bridges} \tilde d_{ia}^k w_i(\mu_i,\mu_a)   
    \frac{\bar v_{ia}}{v_i} \frac{\mu_k}{\mu_i}\frac{\ell_{ia}^k(t)}
    {L_a^2}\right. \nonumber \\ 
    \left. +\sum_{b,\,a\le b}^{\rm zippers} \tilde d_{ab}^i \frac{\bar  
    v_{ab}}{v_i}\frac{(\mu_a+\mu_b-\mu_i)}{\mu_i}\frac{\ell_{ab}^i(t)  
    L_i^2}{L_a^2 L_b^2}\right] \, , \label{v_idtgen}     
  \eq
  where $\tilde d_{ij}^k=\tilde d_{ji}^k$ denotes the efficiency  
  parameter for the process in which strings of type $i$ and $j$  
  interact to produce a type $k$ segment, $\bar v_{ij}$ is the  
  average relative velocity between strings of type $i$ and $j$, 
  whereas $\ell_{ij}^k(t)$ is the average length of links of type  
  $k$, produced by interactions between strings of types $i$ and $j$  
  around time $t$. The first sum in Eq.~(\ref{rho_idtgen}) represents  
  the energy lost from network $i$ due to the length of type $i$  
  strings that coalesced with other types to produce zippers.  Note  
  that no length from the colliding strings is lost when a bridge is  
  produced (see section \ref{bridges}) so the sum is constrained only  
  on interactions of the zipper type.  On the other hand, the second  
  sum models the energy gain of network $i$ due to the production of  
  both bridges and zippers of type $i$.  Similarly, the first sum  
  in Eq.~(\ref{v_idtgen}) describes the slowing down of strings of  
  type $i$ due to their attachment to a newly formed bridge when  
  they interact under the bridge configuration, while the second  
  sum corresponds to the kinetic energy that must be given to  
  zippers of type $i$ to ensure energy conservation.  More 
  general string interactions can be modelled as combinations 
  of the bridge and zipper types, and the relevant weighting 
  can be encoded in the coefficients  $\tilde d_{ij}^k$.  Also, 
  as mentioned above, one can relax the assumption of energy
  conservation at string interactions, to include the possible 
  effect of energy release through particle 
  production~\cite{VinHindSak,MooShelMart}, by modifying some 
  of these coefficients so that string mass-lengths and kinetic 
  energies are not precisely balanced.      

  Working as in sections \ref{basics}, \ref{VOS} we introduce the   
  functions $\gamma_i \equiv L_i/t$ and rewrite Eqs.   
  (\ref{rho_idtgen})-(\ref{v_idtgen}) in the following form 
  \be\label{gamma_idotgen} 
    \frac{\dot\gamma_i}{\gamma_i} = \frac{1}{2t}
    \left[2\beta(1+v_i^2)-2+\frac{\tilde c_i v_i}{\gamma_i} +  
    \sum_{a,k}^{\rm zippers} \tilde d_{ia}^k \bar v_{ia}  
    \frac{1}{\gamma_a^2}\frac{\ell_{ia}^k(t)}{t} - \sum_{b,\,a\le b}   
    \tilde d_{ab}^i \bar v_{ab}  
    \frac{\gamma_i^2}{\gamma_a^2\gamma_b^2}\frac{\ell_{ab}^i(t)}{t} 
    \right]         
  \ee  
  \bq
    \dot v_i = \frac{1-v_i^2}{t}\left[\frac{k_i}{\gamma_i}  
    - 2\beta v_i - \sum_{a,k}^{\rm bridges} \tilde d_{ia}^k   
    w_i(\mu_i,\mu_a) \frac{\bar v_{ia}}{v_i} \frac{\mu_k}{\mu_i}
    \frac{1}{\gamma_a^2} \frac{\ell_{ia}^k(t)}{t}\right. \nonumber \\  
    \left. +\sum_{b,\,a\le b}^{\rm zippers} \tilde d_{ab}^i\frac{\bar  
    v_{ab}}{v_i}\frac{(\mu_a+\mu_b-\mu_i)}{\mu_i}\frac{\gamma_i^2} 
    {\gamma_a^2\gamma_b^2}\frac{\ell_{ab}^i(t)}{t}\right] \, . 
    \label{v_idotgen}
  \eq      
  Eqs.~(\ref{gamma_idotgen})-(\ref{v_idotgen}) can be applied to   
  any given model of multi-string network evolution, specific examples 
  of which will be considered in sections \ref{Z_Nstrings} and \ref{super}.   

  Before moving to these applications, we comment on the size of segments  
  $\ell(t)$ produced in non-abelian interactions which is, in principle,  
  another variable.  One expects this to be a function of the correlation  
  lengths of the colliding strings as well as the string velocities.  In  
  Ref.~\cite{nonint}, $\ell(t)$ was expressed in terms of a constant  
  `zipping velocity' $v_{\rm zip}$, as $\ell(t)\!=\!v_{\rm zip} t$.  In  
  Ref.~\cite{MTVOS} on the other hand, all string types have the same   
  correlation length $L$ and all interactions are of the zipper type,  
  so one can safely assume $\ell(t)\simeq L$.  Here, each string species  
  has its own correlation length so we cannot adopt this approach.  Instead  
  we will consider bridge-type and zipper-type interactions separately and 
  propose physically motivated ansatze. 

  For bridge interactions, the original strings do not lose length but 
  are slowed down due to the tension of the bridge formed between them.   
  Therefore, given the tensions of all strings involved and the velocities  
  and correlation lengths of the colliding segments, one can impose local  
  energy and momentum conservation to find the length of the produced  
  bridge.  For comparable string tensions and correlation lengths, the  
  following approximate formula holds for the length of the produced 
  bridge: 
  \be\label{l_bridge}  
   \ell_{\rm B}\simeq\frac{1}{2\mu_3}(\mu_1 L_1 v_1^2 + \mu_2 L_2 v_2^2) \,.
  \ee 
  This is the formula we used in our bridge toy model, and we will  
  also use it in section \ref{Z_Nstrings}, where we will study the 
  evolution of $Z_N$ strings with similar tensions, through a  
  bridge-type model.  Note that this is the maximum possible length 
  that can be created because all the kinetic energy of the 
  parent strings is lost in stretching the bridge to a length 
  $\ell_{\rm B}$.  By integrating over different string orientations,
  it is clear that the average bridge length will be significantly less 
  than in equation (\ref{l_bridge}), which corresponds to the parameter 
  $\tilde d_{ij}^k$ being significantly smaller than unity 
  $(\tilde d_{ij}^k < 1)$.  

  In the zipper case both of the interacting strings lose the same 
  amount of length (which also equals the length of the produced 
  zipper) and, since string direction changes after correlation 
  length distances on the string, this cannot be larger than the  
  smallest of the two correlation lengths.  We could choose $\ell(t)=  
  {\rm min}(L_1,L_2)$ but this would not be easy to implement in the  
  equations.  Instead, we will take   
  \be\label{l_zipper} 
   \ell=\frac{L_1 L_2}{L_1+L_2} \, , 
  \ee      
  a simple expression which returns a value smaller than, but not far 
  off, the smallest of the two correlation lengths.  This is a good  
  approximation to the smallest correlation length if $L_1$, $L_2$  
  differ by an order of magnitude or more, and it returns half the  
  correlation length for $L_1\!=\!L_2$.  This is not however a problem, 
  as the corresponding term in the evolution equations comes with 
  a coefficient $\tilde d_{ij}^k$, a free parameter of the model    
  in which we can absorb this normalisation.  The important issue   
  is to make sure that $\ell$ has the correct scaling.  We will 
  also use this ansatz when modelling $(p,q)$-strings with zipping 
  interactions in section \ref{super}.     


   For more general interactions, interpolating between the two, 
   the ansatz for $\ell(t)$ will depend on the specific model under 
   study.  Finally, for the collision velocity $\bar v_{ab}$ between 
   strings of type $a$ and $b$ we will take the magnitude of the 
   relative velocity vector ${\bf v}_a - {\bf v}_b$, averaged over 
   directions, namely $\bar v_{ab}=\sqrt{v_a^2+v_b^2}$.

\section{\label{Z_Nstrings}Application to $Z_N$ strings}
 
 Consider a situation in which a continuous, simply connected group $G$  
 is spontaneously broken to $Z_N$. Let $\theta$ (where $0\le \theta \le
 2\pi$) parametrise a closed curve in physical space and denote the
 vacuum at position $\theta$ by $|\theta\rangle$. Then there is a
 group element $g(\theta)\in G$ which maps the vacuum at $0$ to 
 that at $\theta$, that is $|\theta\rangle=g(\theta)|0\rangle$. Now
 consider the unbroken group $Z_N$ at $\theta=0$ with elements $h_i=1, 
 h_2,...,h_N$. The curve under consideration will encircle a string if 
 $g(2\pi)=h_i\ne 1$. One would like to associate a different type of   
 string to each non-trivial group element of the unbroken group.  
 However, the group structure of $Z_N$ imposes relations among various 
 elements, which, under certain circumstances, allow one to identify  
 strings corresponding to different group elements.  

 To see this, consider the cyclic group $Z_N$, which only has one generator  
 $h$ and $N$ elements $h_i=1,h,h^2,...,h^{N-1}$. Acting on $h_N=h^{N-1}$   
 with the generator $h$ gives back the identity, that is   
 $h\cdot h^{N-1}=h^N=1$.  By rewriting this as $h^{N-1}=h^{-1}$ we see   
 that strings corresponding to elements $h_2=h$ and $h_N=h^{N-1}$ have 
 opposite fluxes. In some cases, depending on the details of the symmetry   
 breaking process, such strings and antistrings have been found to be   
 connected by monopoles \cite{HindKib,ArEv}, and the evolution   
 of such monopole-string networks has been discussed in Ref.~\cite{VachVil}.
 However, if the symmetry breaking process does not give rise to 
 monopoles, each string and its corresponding antistring are  
 topologically equivalent \cite{ArEv,AEVV}. In particular,  
 groups $Z_2$ and $Z_3$ give rise to one type of string only, while   
 the lowest order cyclic group which admits more than one distinct    
 strings is $Z_4$.  

 Note however that branching can occur in $Z_3$ networks too \cite{AEVV},  
 even when there is effectively only one type of string (see also  
 Ref.~\cite{Bettencourt}).  To see this, consider a volume element  
 in physical space and assume that an $h_2$ string enters from one  
 side.  Charge conservation requires that the same amount of flux must  
 leave the volume and this can be satisfied, for example, if an $h_2$  
 string also exits the volume from the other side, corresponding to  
 the situation of a single $h_2$ string transversing the volume  
 element~\footnote{Since for $Z_3$ we have $h_2=h_3^{-1}$, this  
 can also be seen as an $h_3$ string entering the volume and  
 joining with the incoming $h_2$ string.}.  However, since  
 $(h_2^{-1})^2=h_3^2=h^4=h=h_2$, flux conservation is also satisfied  
 if two $h_2$ strings enter from the other side, corresponding to three  
 incoming $h_2$ strings joining at a vertex and forming a Y-type junction.   
 But $h_2^{-1}=h_3$ so this can also be seen as an $h_2$ string  
 branching into two (outgoing) $h_3$ strings, two $h_2$ strings joining  
 to an $h_3$, or three $h_3$ strings emerging from a 3-string vertex.   
 From the point of view of string evolution \cite{VachVil}, $h_2$ and  
 $h_3$ strings can be identified and the above configurations are  
 equivalent.

 \subsection{\label{mod_ZN}Modelling $Z_N$ networks} 

\begin{figure}[t]
    \begin{center}
    \includegraphics[height=5cm,width=12cm]{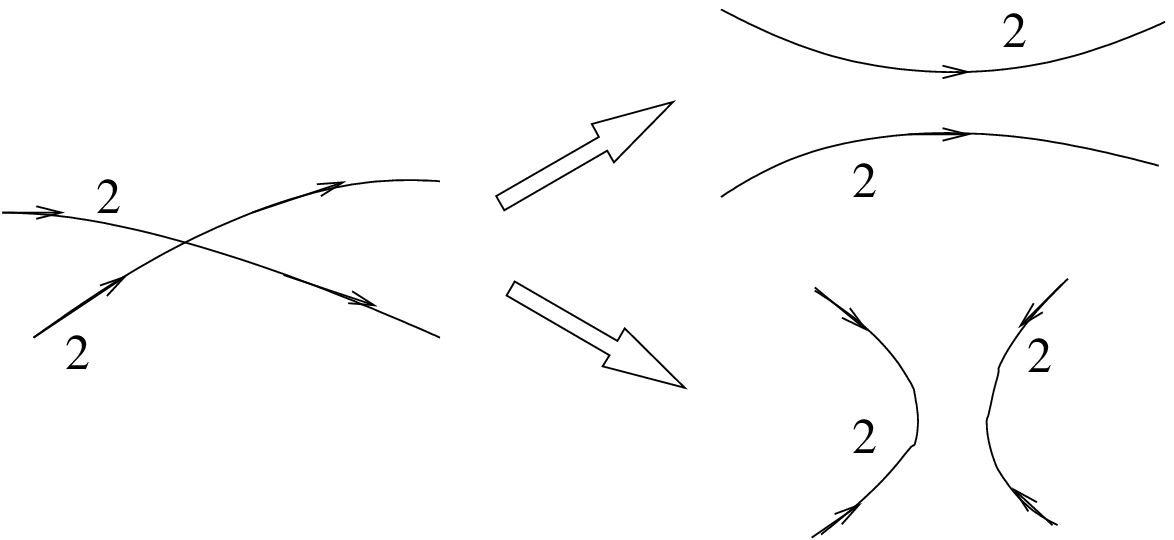}
    \end{center}
    \caption{\label{two_outcomes} For self-conjugate strings,
             reconnection can happen in two distinct ways, as shown.}
   \end{figure}

  In order to apply our string evolution model to a given  
  string network, we need to study the topologically allowed outcomes 
  from interactions between different types of string.  This will  
  allow us to set some of the $\tilde d^i_{jk}$ of equations  
  (\ref{gamma_idotgen})-(\ref{v_idotgen}), namely those  
  corresponding to topologically forbidden interactions, to zero.  
  Consider for example the case of $Z_4$, which has elements  
  $1,h,h^2,h^3$.  There are three types of string corresponding to 
  $h_2=h$, $h_3=h^2$ and $h_4=h^3$.  Since $h\cdot h=h^2$ two strings 
  of type 1 can join to a single type 2 string.  Thus, two intersecting 
  type 1 strings can either exchange partners or form a type 2 string  
  segment (bridge) between them (Fig.~\ref{bridge}).  Similarly,  
  type 1 and type 2 strings can join to a single type 3 string, as  
  $h\cdot h^2=h^3$ (Fig.~\ref{zipper}).  Also, since $h^2\cdot  
  h^2=1$, two type 2 strings can end on a vertex and so, depending  
  on the details of the relevant symmetry breaking, either they  
  are linked by a massive monopole or kink \cite{HindKib}, or they  
  are self-conjugate \cite{ArEv}.  In either case, the crossing of two  
  type 2 strings can have two outcomes, as shown in Fig.~\ref{two_outcomes}. 
  The presence of a massive monopole at the vertex where the strings  
  meet does not directly affect macroscopic string evolution.  Since  
  monopole creation requires energy, such a configuration can result  
  from high energy collisions only.  This can be taken into account in 
  our models by choosing an appropriately small efficiency parameter for 
  such energetically disfavoured processes.  As with type 1 strings, two  
  type 3 segments can join to a single type 2, because $h^3\cdot h^3=h^2$.   
  Finally, $h\cdot h^3=1$ so a type 1 and a type 3 string can also end  
  at a vertex.  Then strings of types 1 and 3 can be seen as the same  
  string, with opposite orientation.   

  In the above example of $Z_4$ strings we see that, although two  
  colliding type 1 strings can form a bridge of type 2 string, 
  intersecting type 2 strings can only exchange partners.  Thus we  
  have $\tilde d^2_{11}\ne 0$, $\tilde d^1_{22}=0$ in  
  (\ref{gamma_idotgen})-(\ref{v_idotgen}).  Also, the intersection  
  between strings of type 1 and 2 can only lead to the formation   
  of a type 1 bridge ($\tilde d^1_{12}=\tilde d^1_{21}\ne 0$) but   
  not type 2 ($\tilde d^2_{12}=\tilde d^2_{21}=0$).  The parameters   
  $\tilde c_i$ and $\tilde d^i_{jk}$ can be grouped in a matrix 
  \be\label{matrix}  
  {\cal M}=
   \left( \begin{array}{cc} (\tilde c_1, \tilde d^2_{11}) & 
                            (\tilde d_{12}^1, 0)             \\ 
                            (\tilde d_{21}^1, 0)          & 
                            (0, \tilde c_2)  
          \end{array} 
   \right) \,. 
  \ee      
  A discussion of the topologically allowed outcomes of string collisions 
  and the corresponding parameter matrices for the cases of $Z_5$, $Z_6$ 
  and $Z_7$ can be found in the appendix.  

  For non-abelian string networks, string-string interactions are governed 
  predominantly by topological, rather than energetic, considerations, 
  except in extreme parameter regimes (e.g. strongly Type I strings).  
  A link is produced between two colliding segments because this is the 
  configuration which conserves charge, not because it minimises energy. 
  Energy gains or losses associated to the colliding strings converting 
  their length to the new string type are only marginal in general.  
  Thus, one expects that such interactions can be approximated by those 
  of the bridge type, in which the interacting strings are not losing 
  length but the link forms and slows them down due its tension.  We 
  will now apply a bridge-type model to the above cases of $Z_N$ strings.  

  \subsubsection{$Z_3$ strings} 
   \begin{figure}[t]
     \begin{center}
     \includegraphics[height=2.2in,width=2.1in]{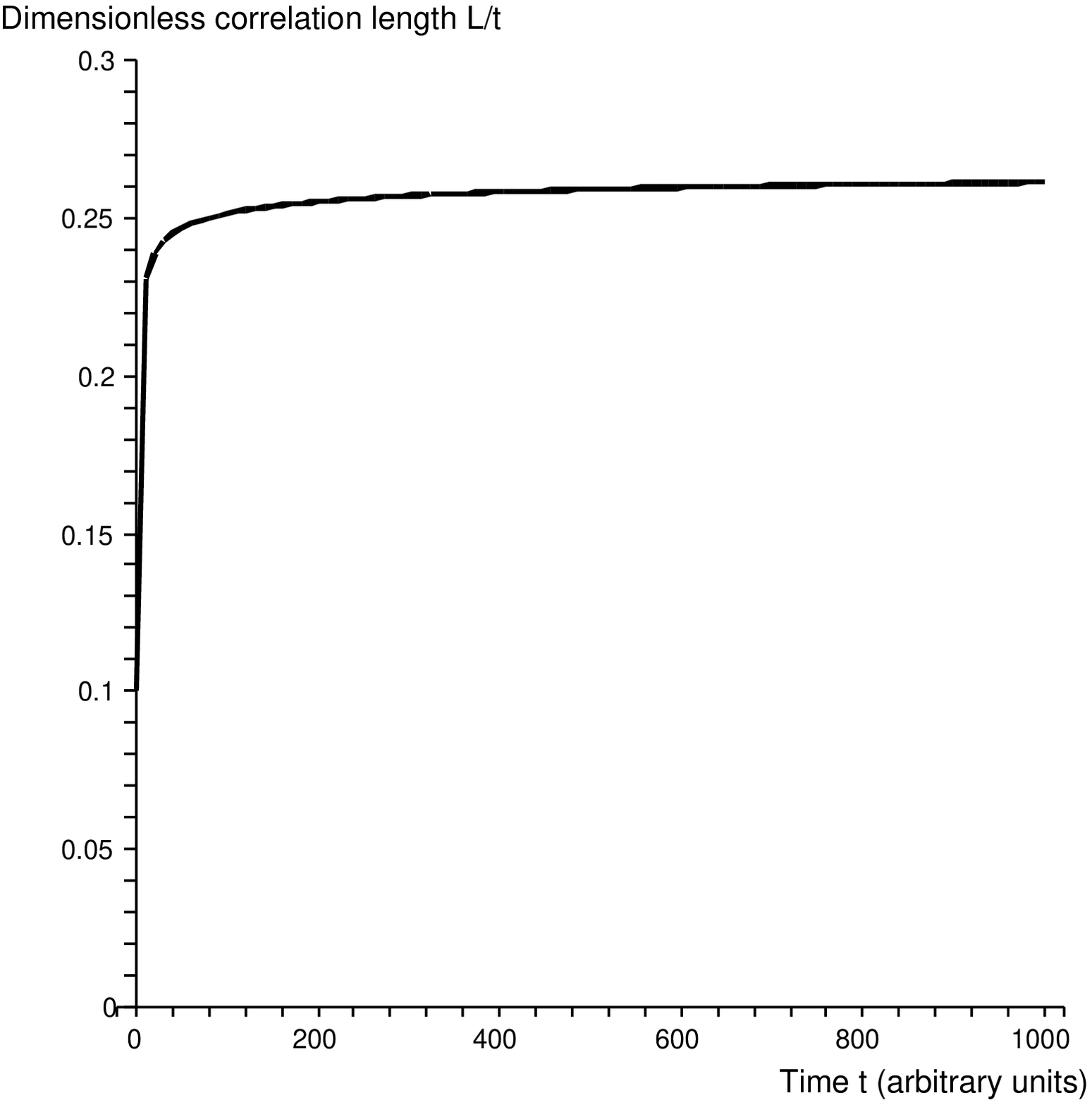}
     \includegraphics[height=2.2in,width=2.1in]{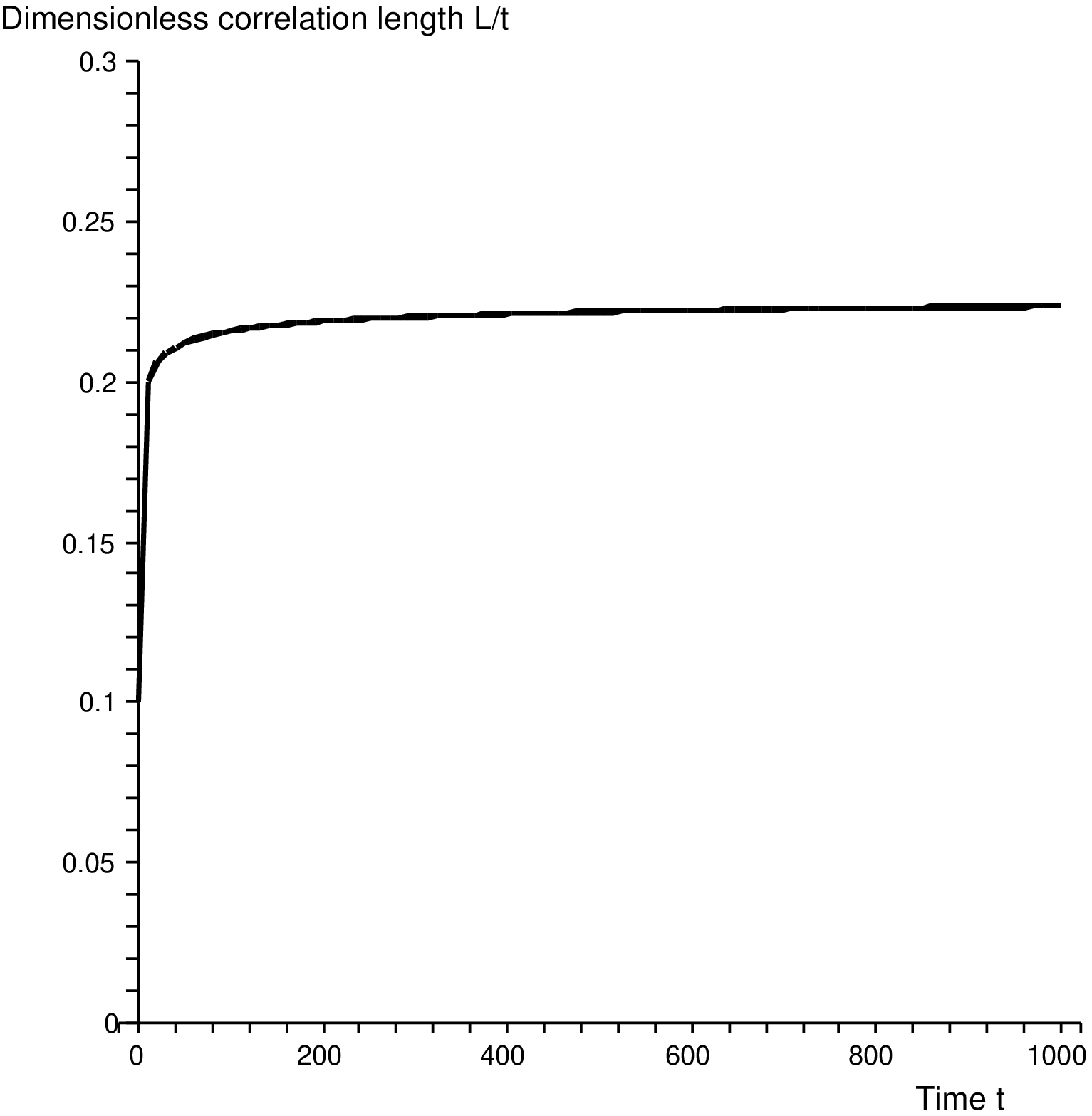}
     \includegraphics[height=2.2in,width=2.1in]{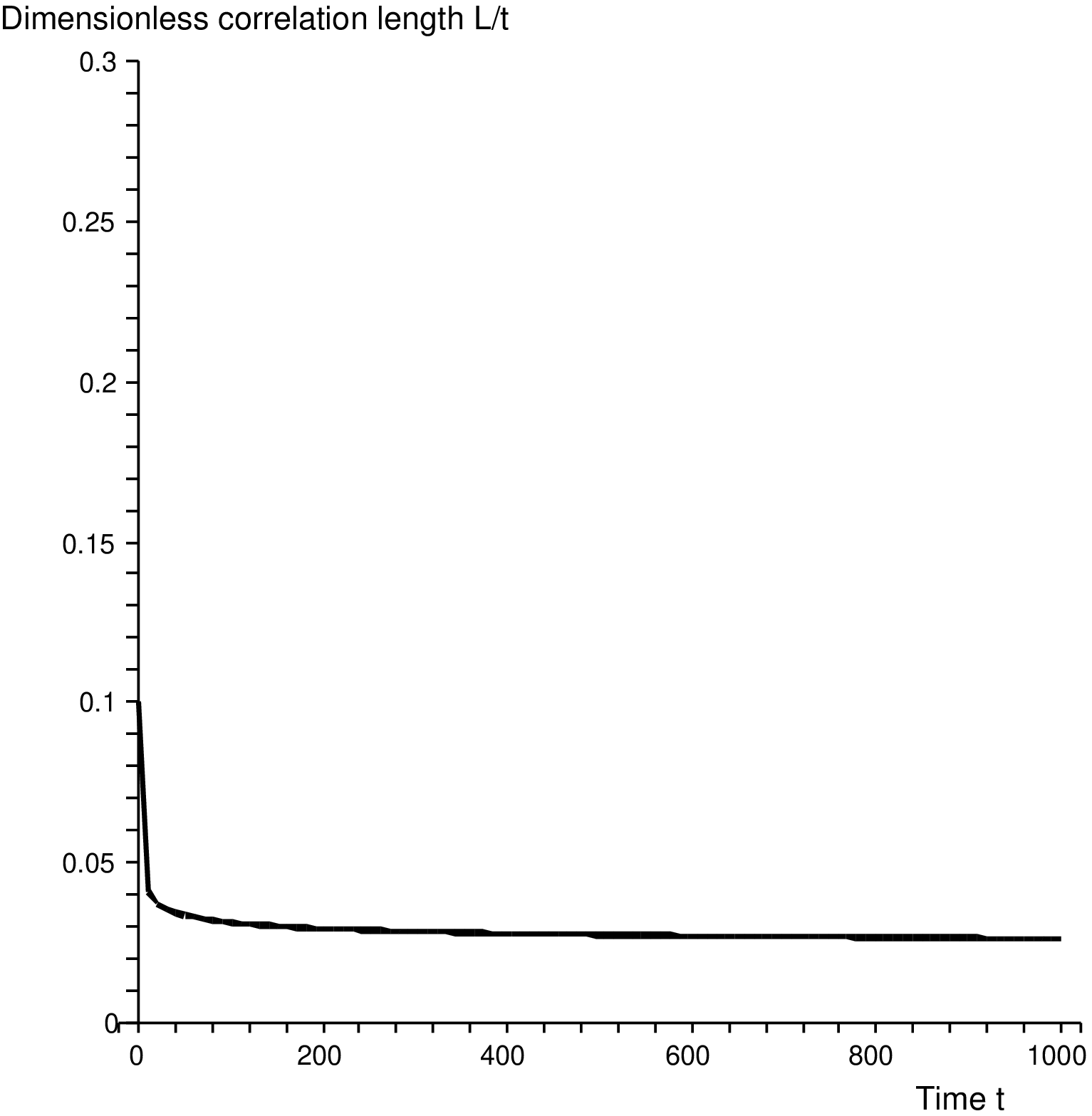}
     \includegraphics[height=2.2in,width=2.1in]{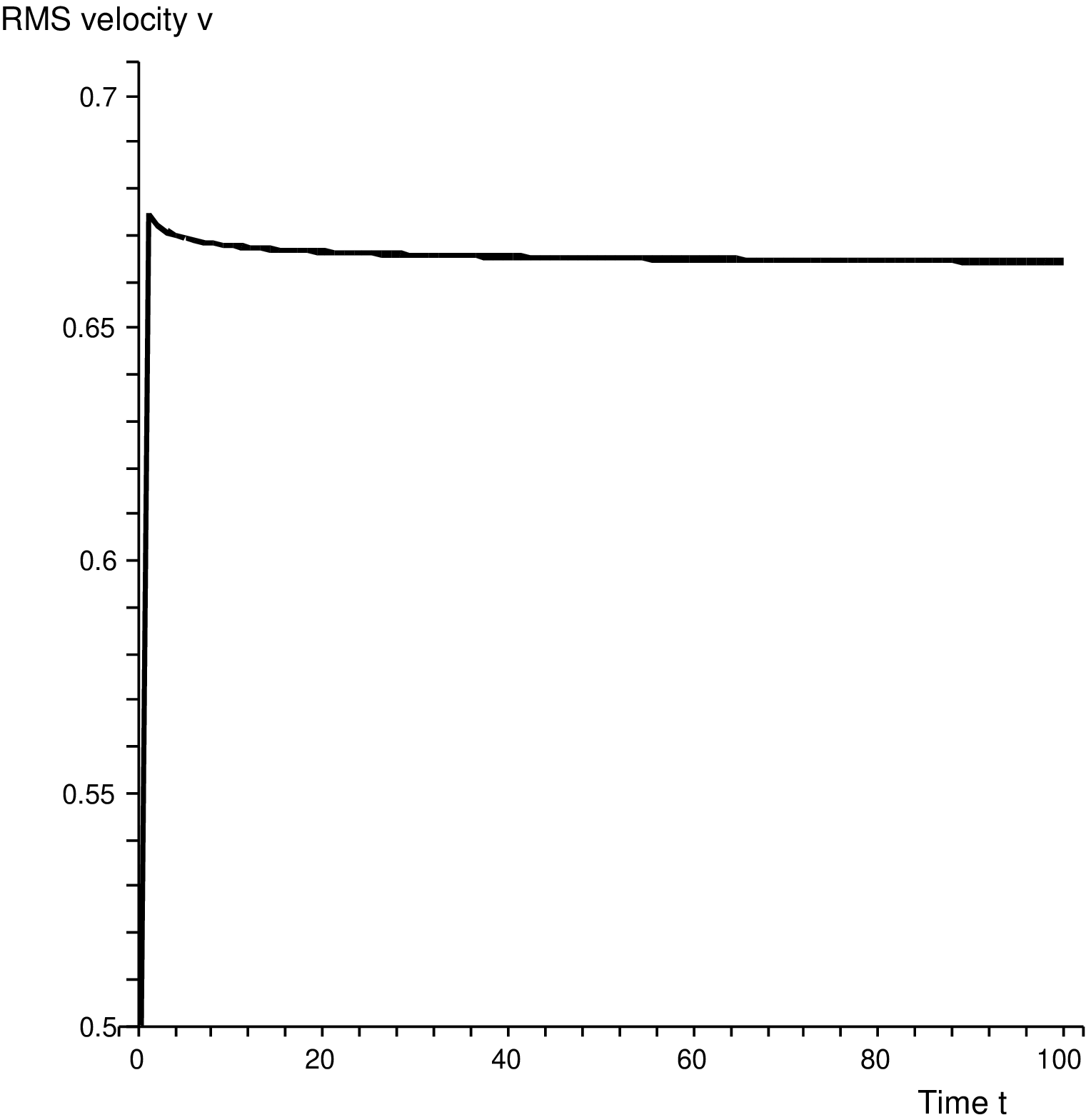}
     \includegraphics[height=2.2in,width=2.1in]{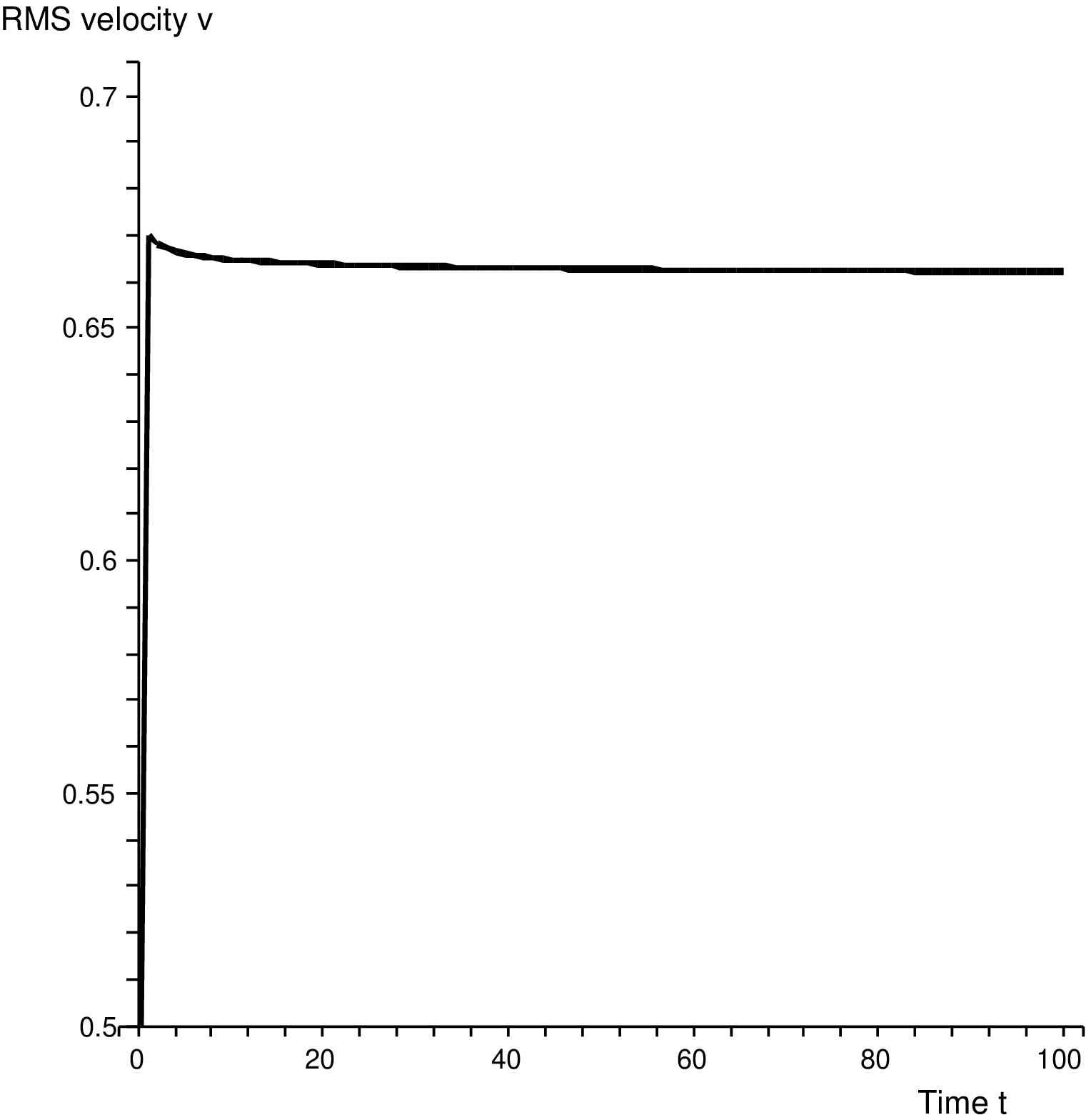}
     \includegraphics[height=2.2in,width=2.1in]{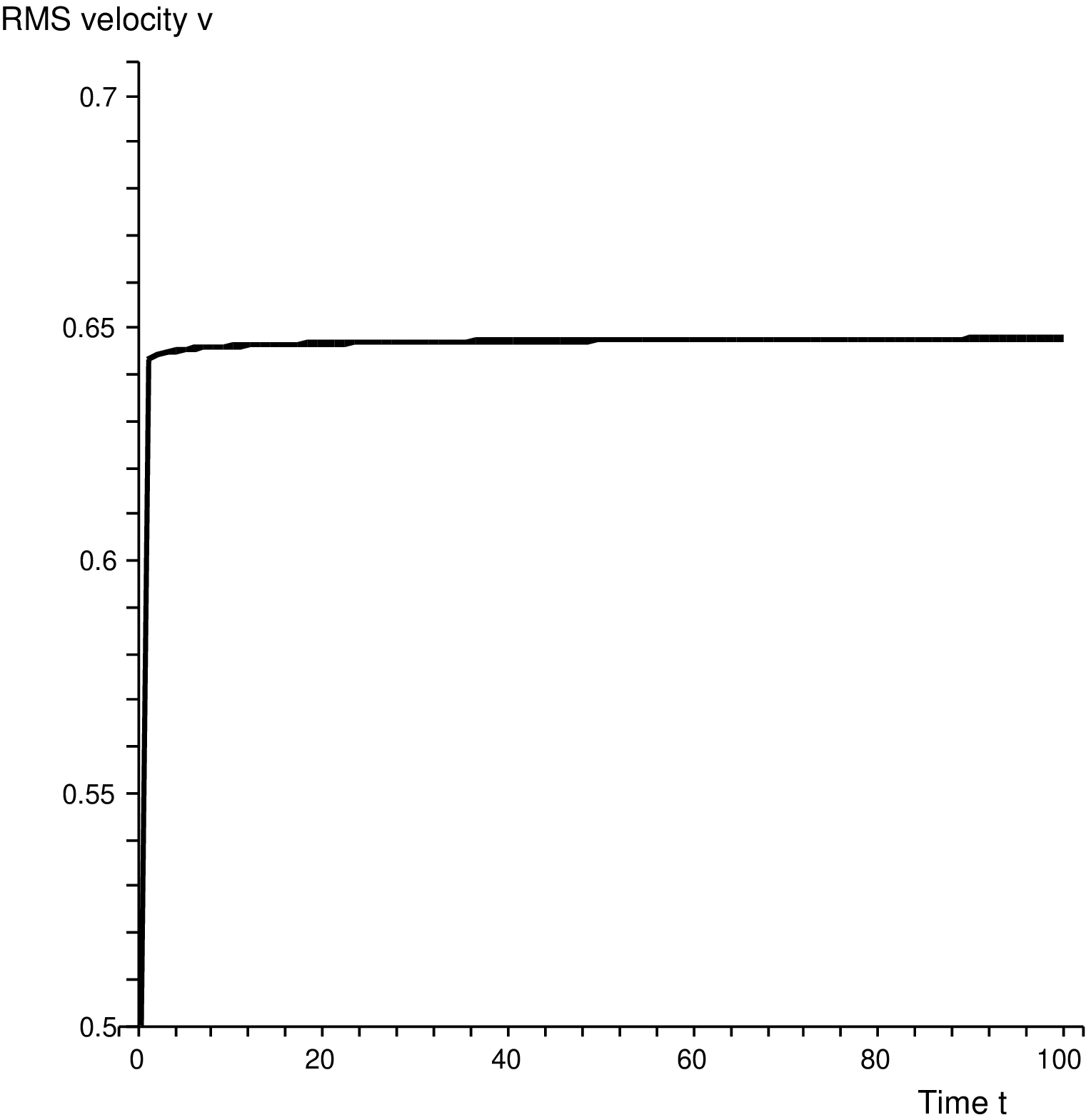}
     \end{center}
     \caption{\label{Z3} Evolution of $\gamma$ and $v$ for a $Z3$
              network, where there is only one type of string,
              for $\tilde d=0$, $0.1\tilde c$ and $0.65\tilde c$.
              The string network reaches scaling with smaller
              $\gamma$ as $\tilde d$ increases.  This corresponds
              to a larger string energy density as more Y-type
              junctions are being produced.}
   \end{figure}
  The simplest case allowing the formation of Y-type junctions is $Z_3$,  
  where there is only one type of string and the collision between two  
  string segments can either lead to reconnection or to the production  
  of a bridge.  The corresponding VOS model describing the evolution of  
  such a network is given by the equations (see  
  Eqs.~(\ref{gamma_idotgen})-(\ref{v_idotgen})) 
  \be\label{gammadotZ3}
   \frac{\dot\gamma}{\gamma} = \frac{1}{2t}
   \left[2\beta(1+v^2)-2 + \frac{\tilde c v}{\gamma} -
   \tilde d \frac{\bar v}{\gamma^2}\frac{\ell(t)}{t}
   \right]
  \ee
  \be\label{vdotZ3}
    \dot v = \frac{1}{t}\left[(1-v^2)\left(\frac{k}{\gamma}
    - 2\beta v - \tilde d \frac{\bar v}{v} \frac{1}{\gamma^2}
    \frac{\ell(t)}{t} \right) \right]  , 
  \ee  
  where $\bar v=\sqrt{2}v$ is the average relative velocity of strings  
  and all other parameters/variables are as described in the text.  
  Solving this system with $\tilde c=0.23$ we find (Fig.~\ref{Z3}) scaling 
  solutions for $\tilde d \lesssim 0.7\tilde c$.  The larger $\tilde  
  d$ is, the more entangled the network becomes and the scaling solution   
  moves from its `abelian' value to one with higher string density.   
  For $\tilde d \gtrsim 0.7 \tilde c$ the network becomes so dense   
  that the bridge production term dominates, driving the system to  
  an even denser state and thus spoiling scaling.  In view of the   
  discussion of last paragraph of \ref{bridges} this happens in the   
  case of $Z_3$ because both the interacting strings and the   
  produced bridge are of the same type, so the dependence of the   
  bridge production term on $\gamma$ changes from $\gamma^2$ to 
  $\gamma^{-2}$.  Then if this term gets to dominate, the string  
  density increases ($\gamma$ decreases) and the term becomes even  
  stronger, so that scaling cannot be achieved.       
  However, since intercommuting is energetically favourable compared 
  to bridge production, one expects $\tilde d < \tilde c$.  Then 
  the formation of Y-type junctions increases the string density  
  (Fig.~\ref{Z3}) but does not spoil the scaling property of the 
  network.  
  \begin{figure}[h!]
    \includegraphics[height=2.7in,width=2.9in]{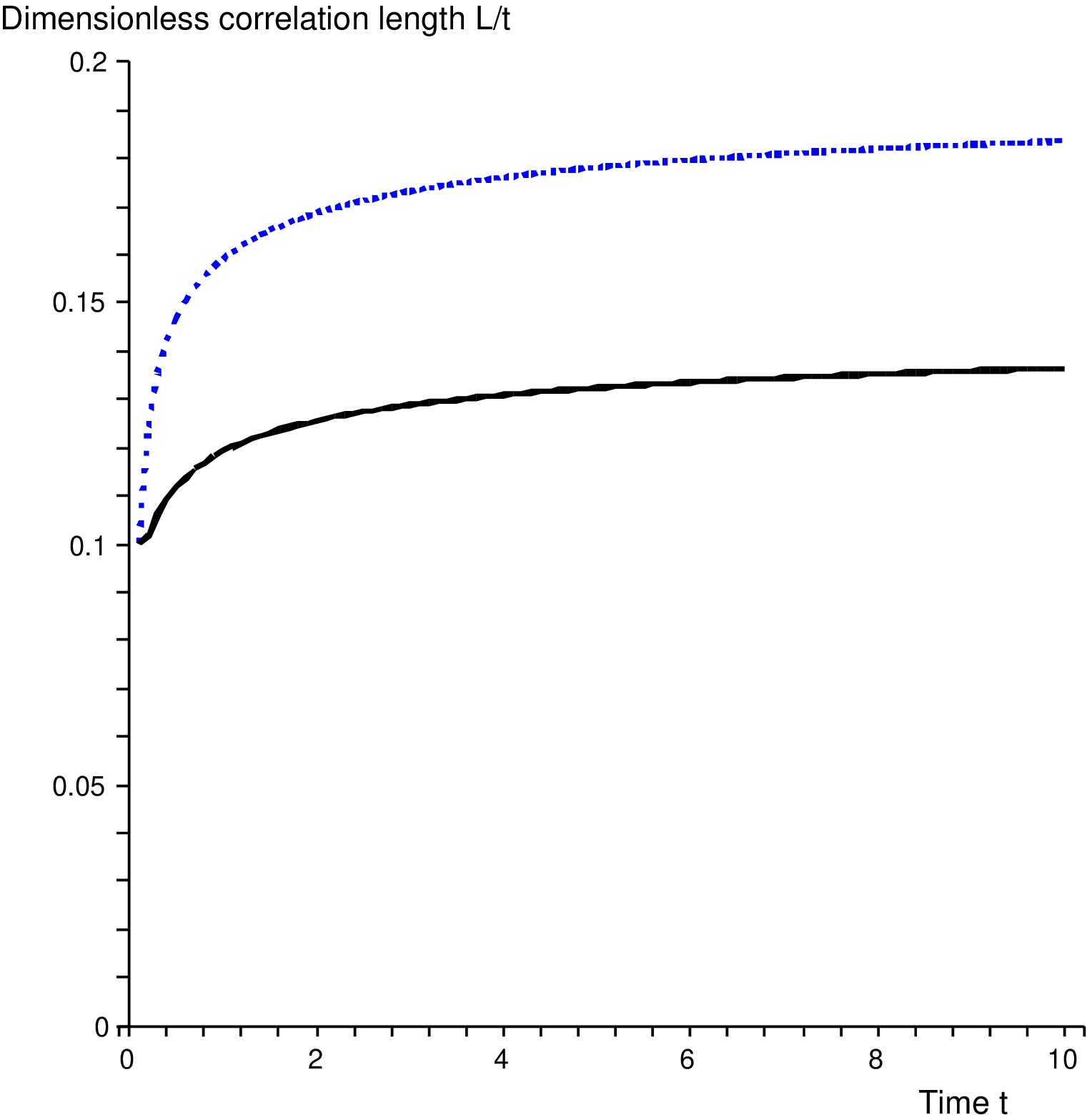}
    \includegraphics[height=2.7in,width=2.9in]{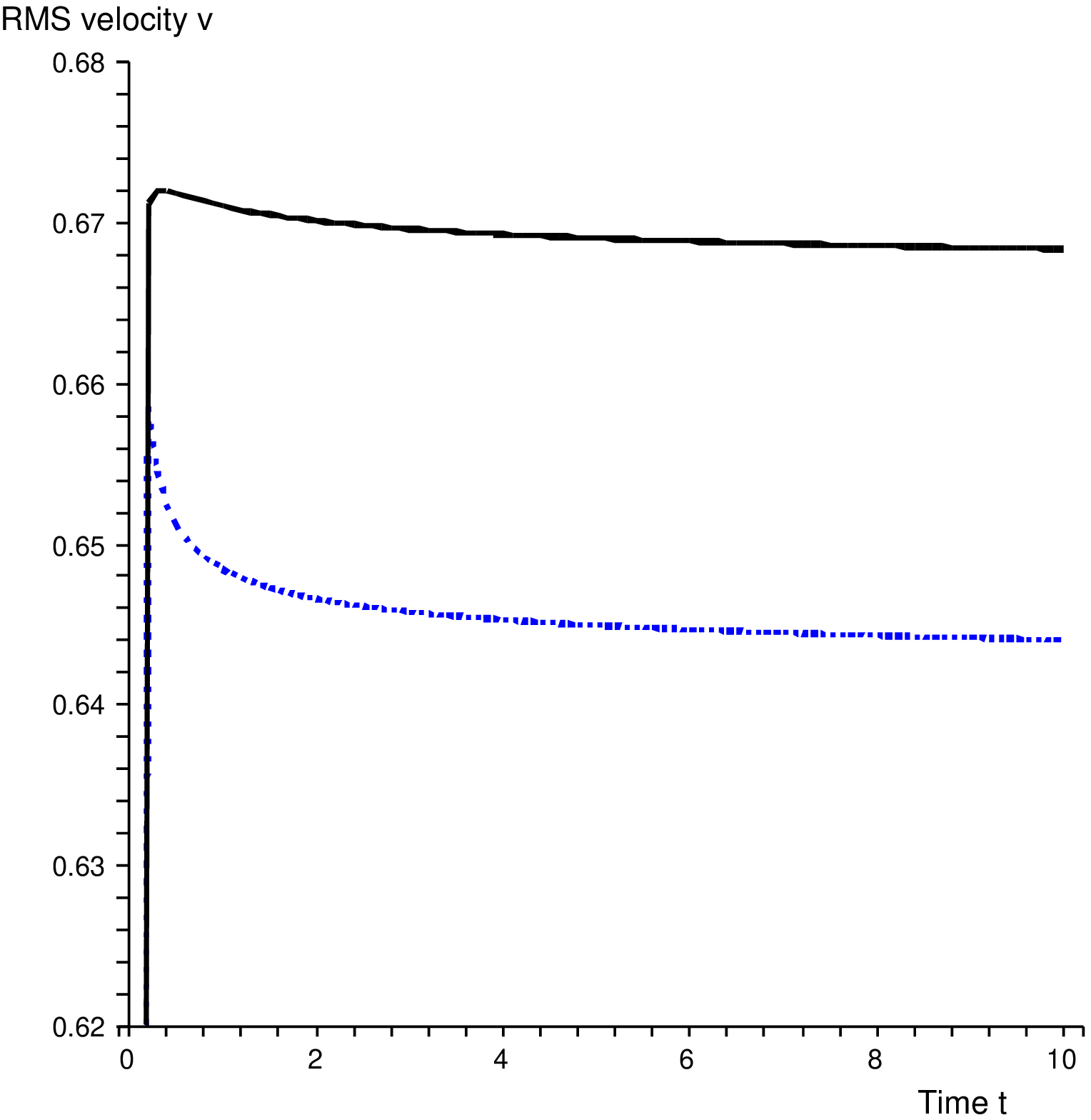}
    \caption{\label{Z4} Evolution of $\gamma$ and $v$ for
             the two components of a $Z4$ network, for $\tilde c_1
             =\tilde c_2=0.23$, $\tilde d_{11}^2=0.05\tilde c_1$ and
             $\tilde d_{12}^1=\tilde c_1$.  Type 1 strings are
             represented by a solid black  line and type 2 by a
             dotted blue one.  We observe that type 1 strings are
             more abundant, since, unlike type 2, they can also be
             produced by interactions between type 1 and type 2
             strings.}
   \end{figure}

  \subsubsection{$Z_4$ strings}
  For $Z_4$ there are two types of string, say type 1 and type 2,  
  and the corresponding parameter matrix is given by (\ref{matrix}).  
  A $Z_4$ network can therefore be described by two copies of  
  equations (\ref{gamma_idotgen})-(\ref{v_idotgen}) with non-zero  
  parameters $\tilde c_1$, $\tilde c_2$, $\tilde d_{11}^2$ and  
  $\tilde d_{12}^1=\tilde d_{21}^1$ (and with all zipper terms  
  set to zero).  Again scaling solutions can be found when the  
  bridge production terms do not dominate.  For type 1 strings, 
  since reconnection is energetically favourable compared to  
  bridge production, we will assume $\tilde d_{11}^2<\tilde c_1$.   
  For collisions between type 1 and 2 strings, however, reconnection  
  is not possible and the production of a type 1 bridge is the 
  only option, so that $\tilde d_{12}^1$ can be large.  One then  
  finds scaling as long as $\tilde d_{12}^1$ does not exceed a  
  critical value, which depends on the chosen $\tilde d_{11}^2$. 
  For example choosing $\tilde c_1=\tilde c_2=0.23$, $\tilde d_{11}^2 
  =0.05\tilde c_1$ (Fig.~\ref{Z4}) we find scaling, for both string 
  types, for $\tilde d_{12}^1\lesssim 1.5 \tilde c_1$.  

  \subsubsection{$Z_5$ strings} 
\begin{figure}[h]
    \includegraphics[height=2.7in,width=2.9in]{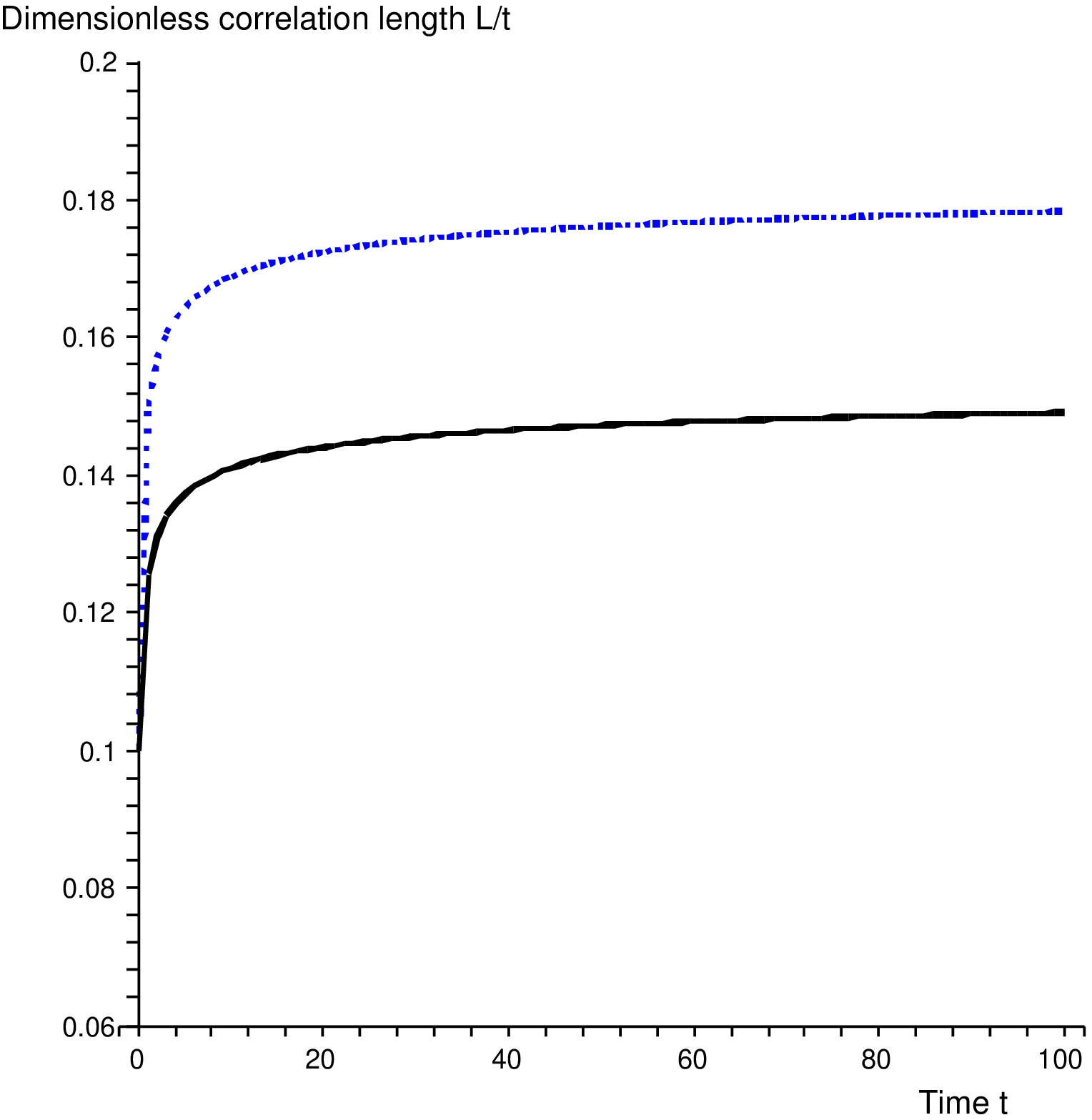}
    \includegraphics[height=2.7in,width=2.9in]{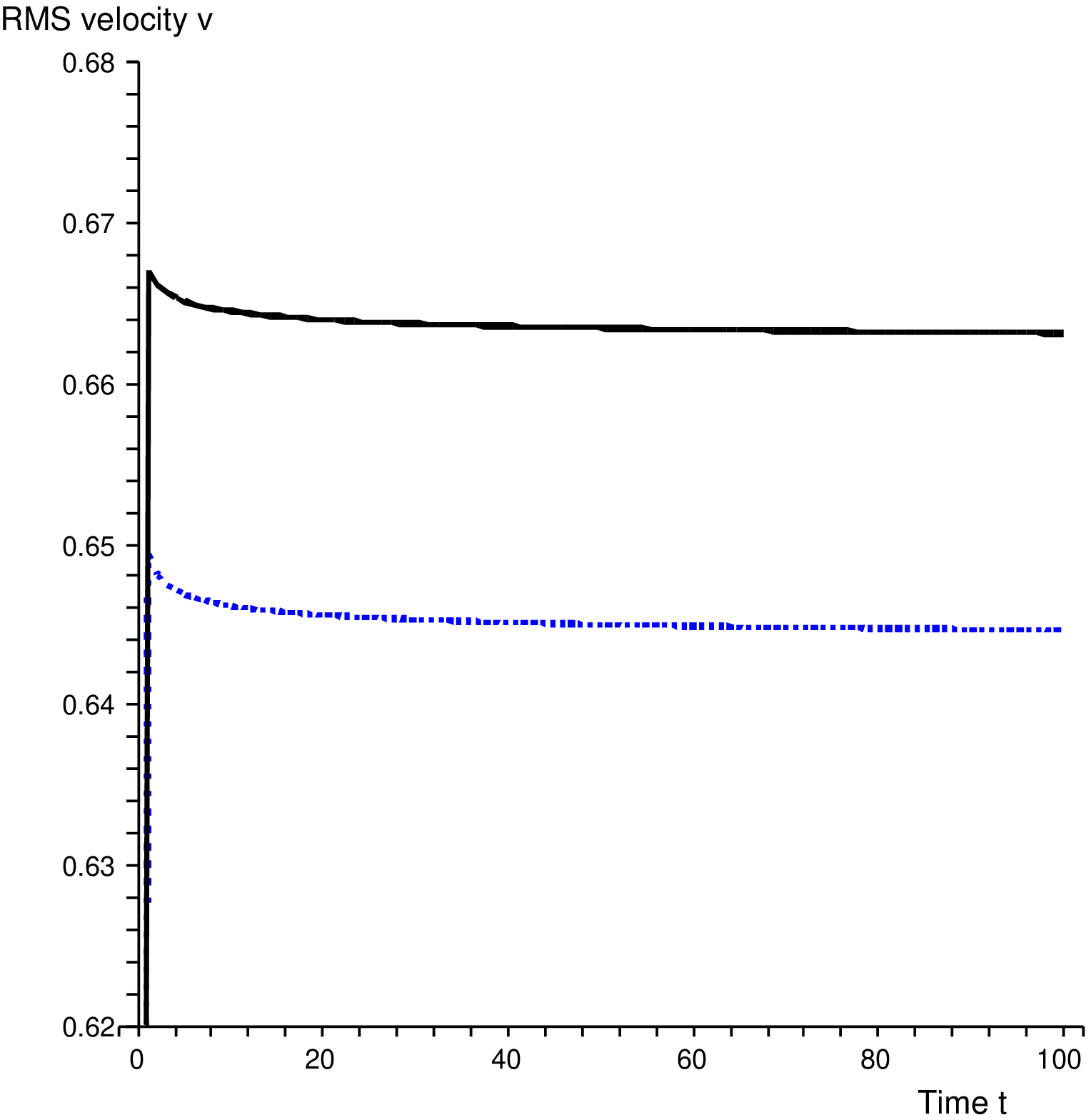}
    \caption{\label{Z5} Evolution of $\gamma$ and $v$ for
             the two components of a $Z5$ network. In this example  
             we have chosen $\tilde c_1=0.23$, $\tilde c_2=0.3$,  
             $\tilde d_{11}^2=0.05\tilde c_1$, $\tilde d_{22}^1=  
             0.05\tilde c_2$, $\tilde d_{12}^1=0.6\tilde c_1$ 
             and $\tilde d_{12}^2=0.4\tilde c_1$.  Choosing  
             $\tilde c_1=\tilde c_2$ and  $\tilde d_{12}^1= 
             \tilde d_{12}^2$ instead leads to identical  
             evolutions for the two string types, as the  
             corresponding evolution equations are symmetric.}
   \end{figure}

  $Z_5$ networks also consist of two types of string but with  
  different interaction rules (see appendix).  The parameter  
  matrix is given by (\ref{matrix_Z5}) and the relevant VOS model  
  arises by the appropriate adaptation of two copies of equations 
  (\ref{gamma_idotgen})-(\ref{v_idotgen}).  Again, both string types 
  achieve scaling when $\tilde d_{ij}^k$ are not larger than $\tilde  
  c$ (Fig.~\ref{Z5}).   
  
  \subsubsection{$Z_6$ strings} 
\begin{figure}[h]
    \includegraphics[height=2.8in,width=2.9in]{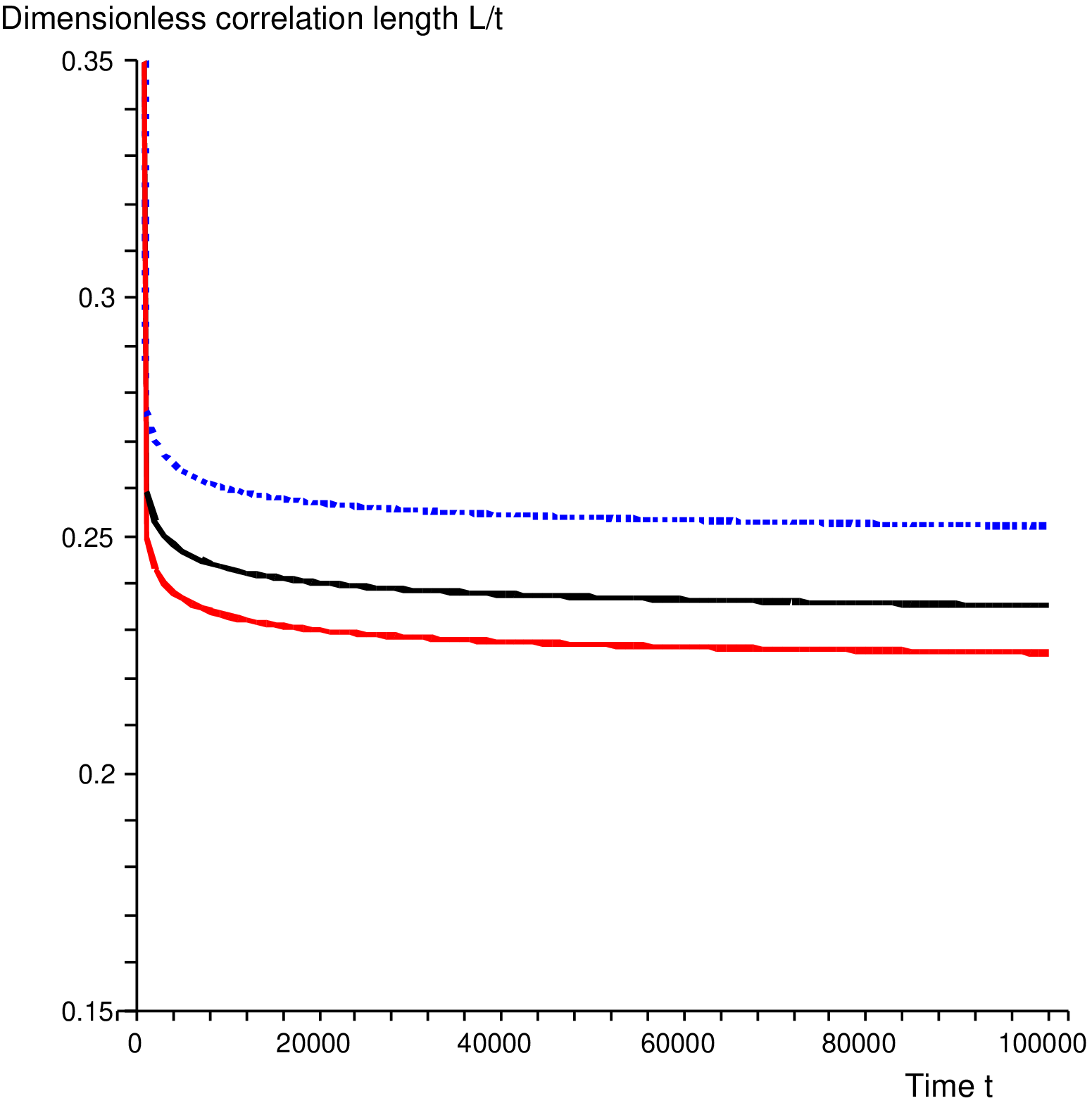}
    \includegraphics[height=2.8in,width=2.9in]{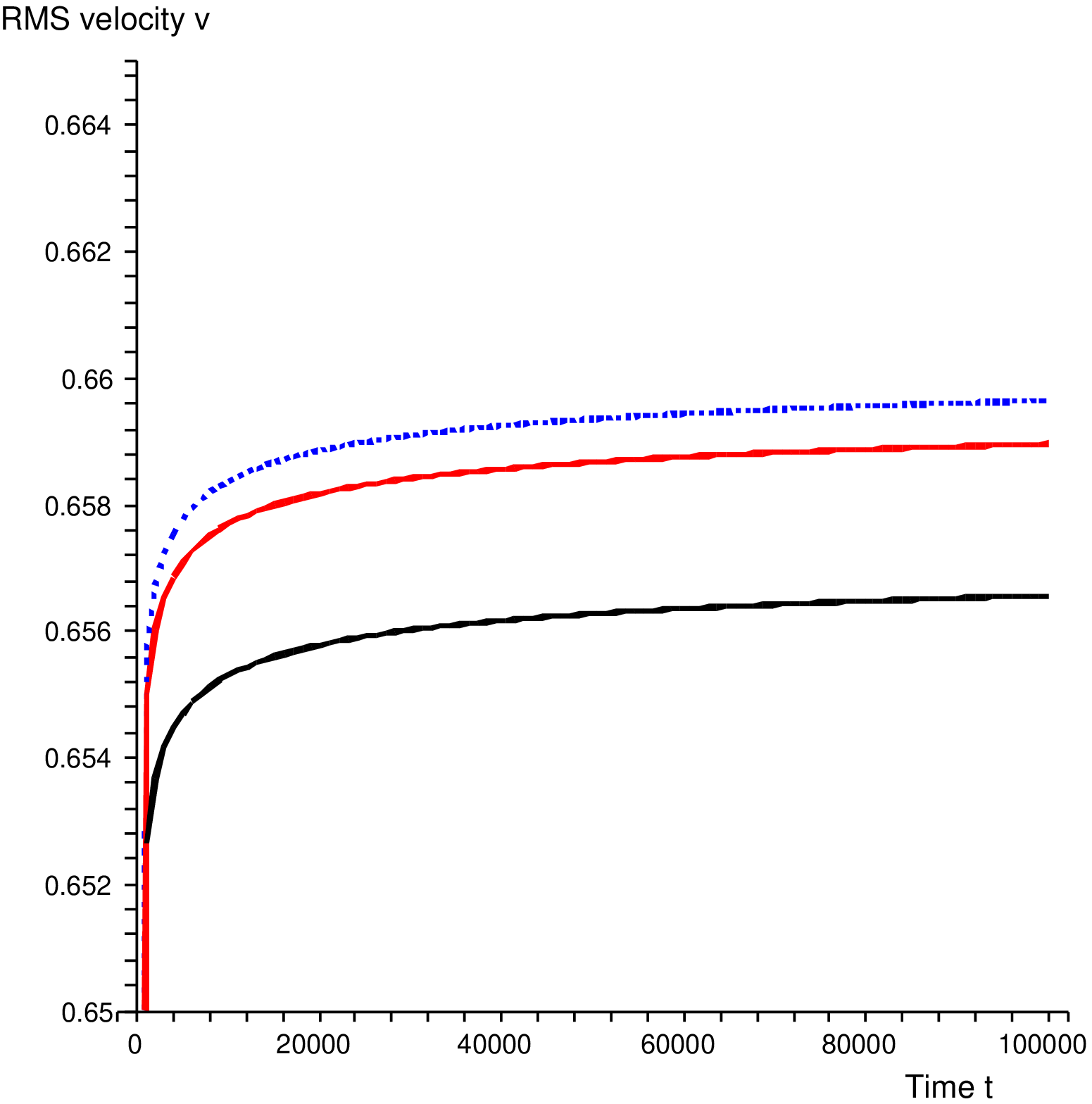}
    \caption{\label{Z6} Evolution of $\gamma$ and $v$ for
             the three components of a $Z6$ network. Here, we 
             have chosen $\tilde c_1=\tilde c_2=\tilde c_3=0.23$,  
             $\tilde d_{11}^2=\tilde d_{22}^2=0.05\tilde c_1$ and
             $\tilde d_{12}^1=\tilde d_{12}^3=\tilde d_{13}^2=0.1 
             \tilde c_1$.  Type 1 strings are represented by a solid  
             black line, type 2 by a solid red, and type 3 by a dotted 
             blue line.}
   \end{figure}

  In the case of $Z_6$ there are three distinct types of strings 
  (see appendix) with parameter matrix (\ref{matrix_Z6}).  The system 
  is modelled by three copies of (\ref{gamma_idotgen})-(\ref{v_idotgen}).  
  All three string types reach scaling for small enough  
  $\tilde d_{ij}^k$'s (Fig.~\ref{Z6}).  
 
  \subsubsection{$Z_7$ strings}
\begin{figure}[h]
    \includegraphics[height=2.8in,width=2.9in]{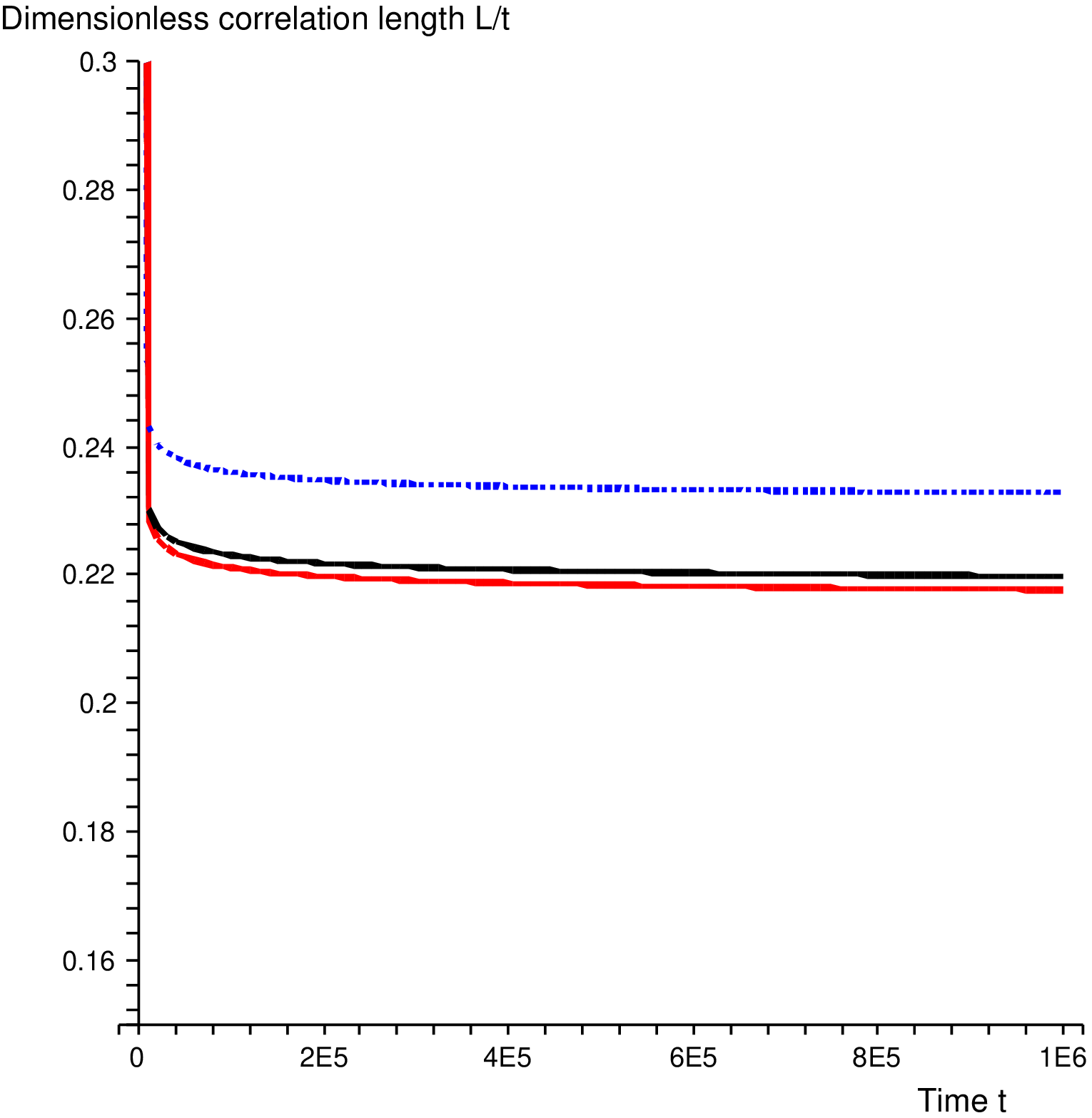}
    \includegraphics[height=2.8in,width=2.9in]{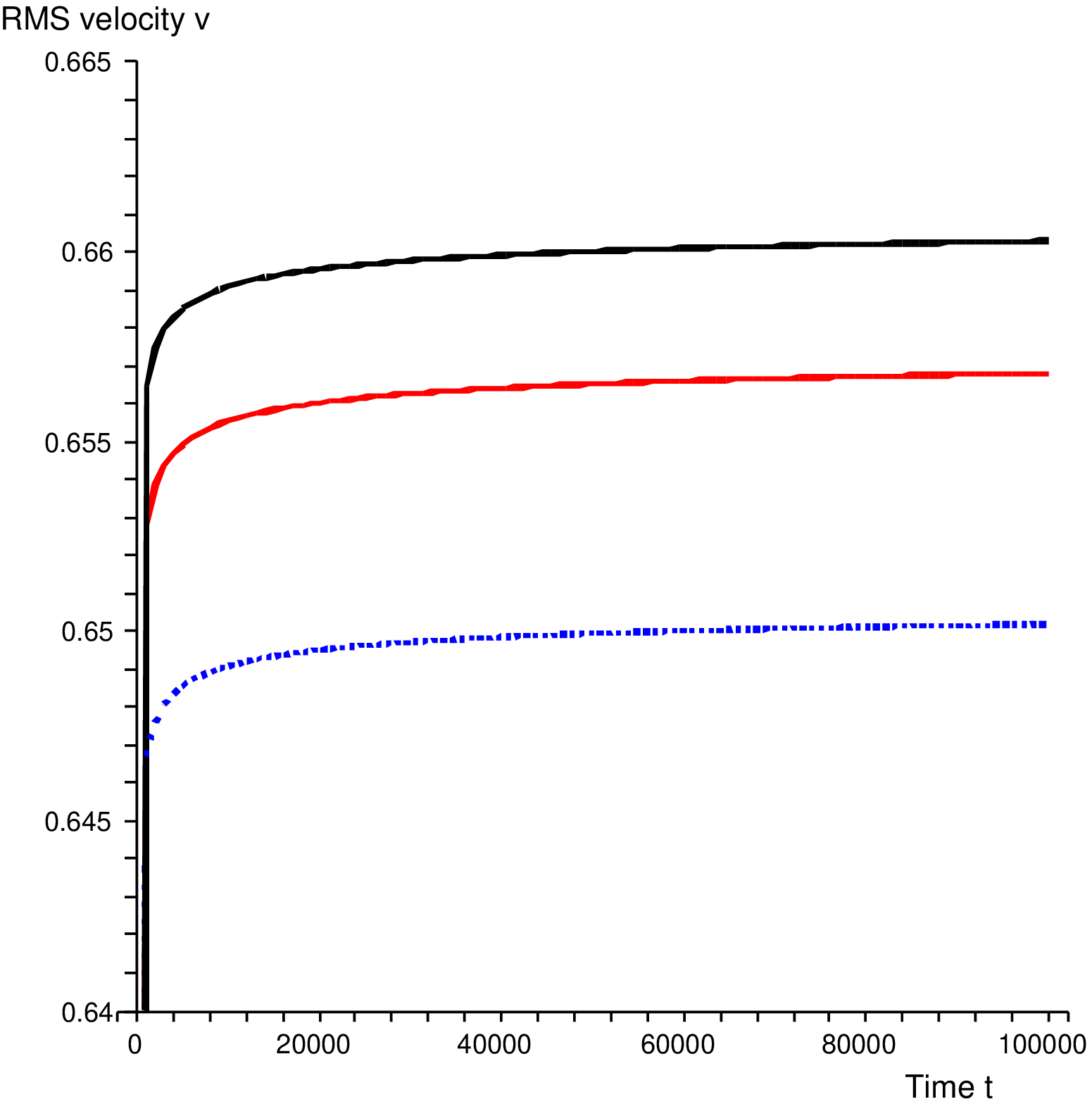}
    \caption{\label{Z7} Evolution of $\gamma$ and $v$ for
             the three components of a $Z7$ network. We have 
             chosen $\tilde c_1=\tilde c_2=\tilde c_3=0.23$,
             $\tilde d_{11}^2=0.05\tilde c_1$ and $\tilde d_{12}^1 
             =\tilde d_{12}^3=\tilde d_{13}^2=\tilde d_{13}^3= 
             \tilde d_{22}^3=\tilde d_{23}^1=\tilde d_{23}^2= 
             \tilde d_{33}^1=0.1\tilde c_1$.  As before, type 1 
             strings are represented by a black solid line,   
             type 2 by a red solid and type 3 by a blue dotted  
             line.}
   \end{figure}

  Here there are also three string types with parameter matrix  
  (\ref{matrix_Z7}).  The relevant scaling solutions are shown in 
  Fig.~\ref{Z7}. 


\section{\label{super}The case of cosmic superstrings}

 In this section we apply our model to cosmic superstrings, typically  
 produced at the end of brane inflation \cite{DvalTye,DvalShafSolg,BMNQRZ, 
 Garc-Bell,KKLMMT}.  In this picture, a brane-antibrane pair are moving  
 towards each other under their attractive interaction, and in doing so  
 they give rise to an inflationary phase.  Inflation ends when the branes   
 collide and annihilate, leaving behind a network of D- and F-strings  
 \cite{BMNQRZ,SarTye,JoStoTye2,DvalVil,PolchProb}.  String interactions  
 can lead to the formation of bound states between $p$ F-strings and  
 $q$ D-strings, referred to as $(p,q)$-strings, where $p,q$ are  
 coprime~\footnote{If $p,q$ are not relatively prime, the string is  
 just a collection of lighter coprime strings.}.  This situation  
 corresponds more closely to interactions of the zipper type, rather 
 than bridge production, as the strings coalesce along their own length 
 in forming a bound state.  Also, unlike $Z_N$ strings, where different 
 string types could have the same tension, here each type of string has  
 different tension.  F-strings, being perturbative objects have   
 tensions proportional to the square root of the string coupling,  
 namely   
 \be\label{Ftension} 
  \mu_F=\mu_0 \sqrt{g_s} \,\,, 
 \ee
 while D-strings are non-perturbative and have 
 \be\label{Dtension} 
  \mu_D=\mu_0 / \sqrt{g_s} \,\,.
 \ee
 In flat space\footnote{For the corresponding formula in warped space 
 see Ref.~\cite{FirLebTye}.  String evolution in this context has been 
 studied in \cite{LebWy,warped}.} the tension of $(p,q)$ strings is 
 given by  
 \be\label{pqtension} 
  \mu_{(p,q)}=\mu_F \sqrt{p^2+q^2/g_s^2} \,\,.   
 \ee

 \subsection{\label{mod_super}Modelling superstring networks} 

\begin{figure}[b]
    \includegraphics[height=7cm,width=12cm]{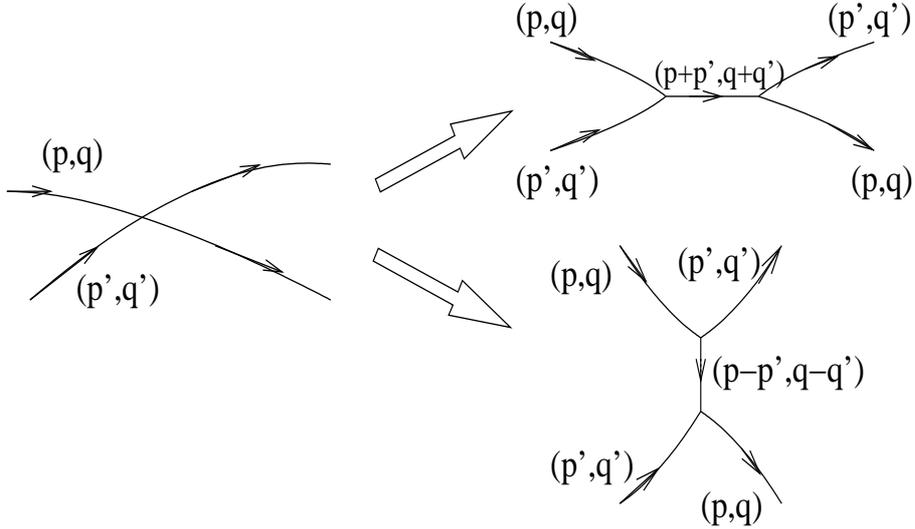}
    \caption{\label{p_plusminus_q} A $(p,q)$ and a $(p',q')$ strings,   
             can bind together in two distinct ways, forming either 
             a $(p+p',q+q')$ or a $(p-p',q-q')$ zipper.}
  \end{figure}

 To model cosmic superstring networks we consider two types of  
 \emph{elementary} string, type 1 with tension $\mu_1=\mu_0 \sqrt{g_s}$   
 (corresponding to the F-string) and type 2, tension $\mu_2=\mu_0 /  
 \sqrt{g_s}$ (the D-string).  When two elementary strings of the same  
 type collide, they may pass through one another or reconnect by   
 exchange of partners.  However, a type 1 and a type 2 string can  
 bind together to form a $(1,\pm 1)$ string segment.  In general, when  
 a $(p,q)$ and a $(p',q')$ string collide they can bind in two ways,  
 depending on the angle of collision \cite{MTVOS,PolchProb,PolchStab},  
 forming  either a $(p+p',q+q')$ or a $(p-p',q-q')$ segment  
 (Fig.~\ref{p_plusminus_q}), where without loss of generality we have   
 taken $p>p'$.   
 Given that binding has taken place, the probability  
 that the additive/subtractive process has occurred is given by \cite{MTVOS} 
 \be\label{Paddsubtr} 
  P_{(p,q),(p',q')}^\pm = \frac{1}{2}\left( 1\mp\left( \frac{pp'g_s^2 
  +qq'}{(p^2g_s^2+q^2)^{1/2}\,(p'^2g_s^2+q'^2)^{1/2}} \right) \right) , 
 \ee
 where we have assumed that the RR scalar is zero.  Assuming that the 
 zipper type interaction is a good approximation to this binding, we can 
 model such a network by equations (\ref{gamma_idotgen})-(\ref{v_idotgen}) 
 with zero bridge terms and  
 \be\label{d_super}
  \tilde d_{(p,q),(p',q')}^{(p\pm p',q\pm q')} = 
  \tilde d_{(p,q),(p',q')} P^\pm ,  
 \ee       
 with $\tilde d_{(p,q),(p',q')}$'s in the range found in  
 Ref.~\cite{PolchProb}. 
 
\begin{figure}[h]
  \begin{center}
    \includegraphics[height=3.1in,width=3.2in]{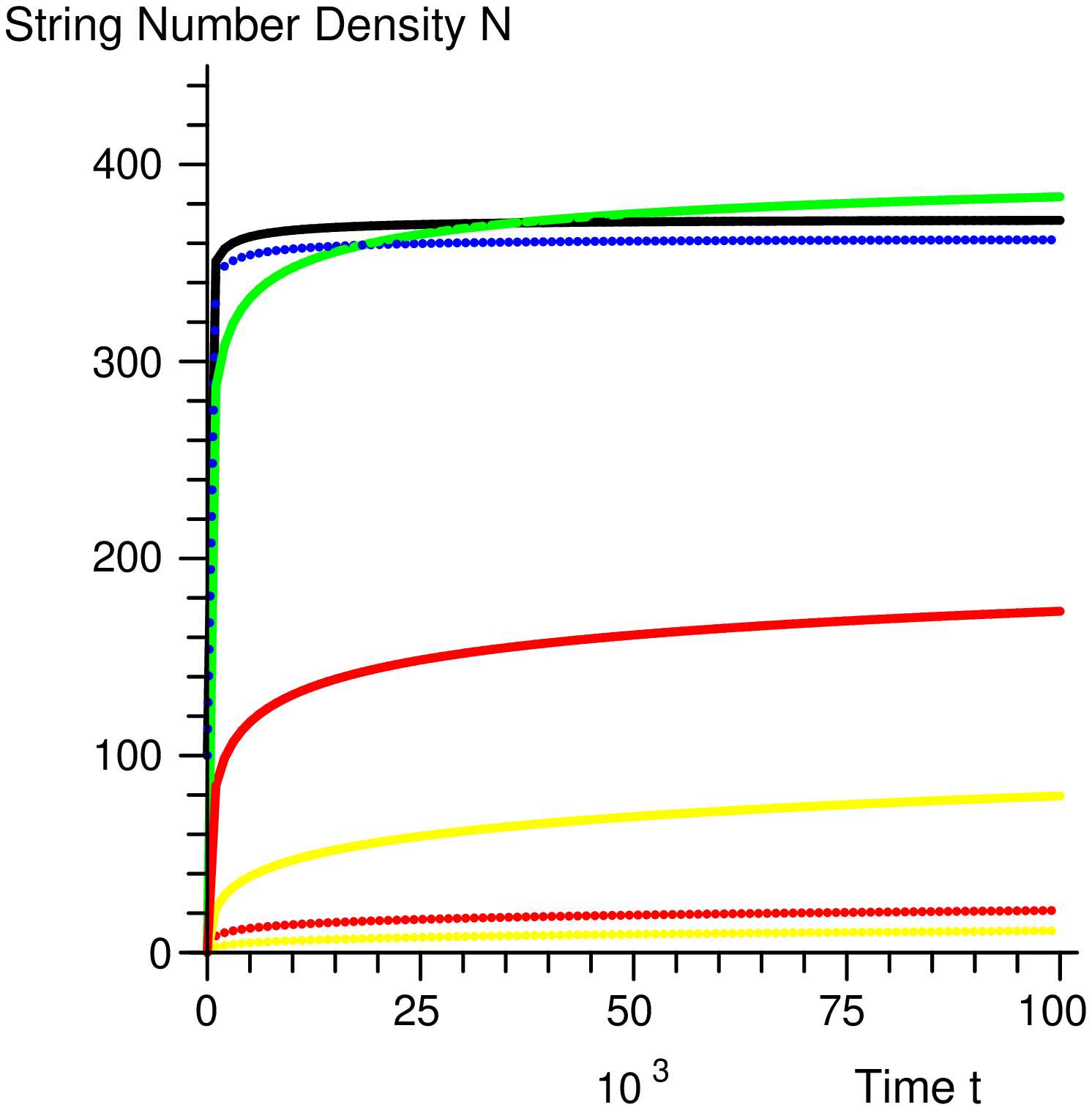}
    \includegraphics[height=3.1in,width=3.2in]{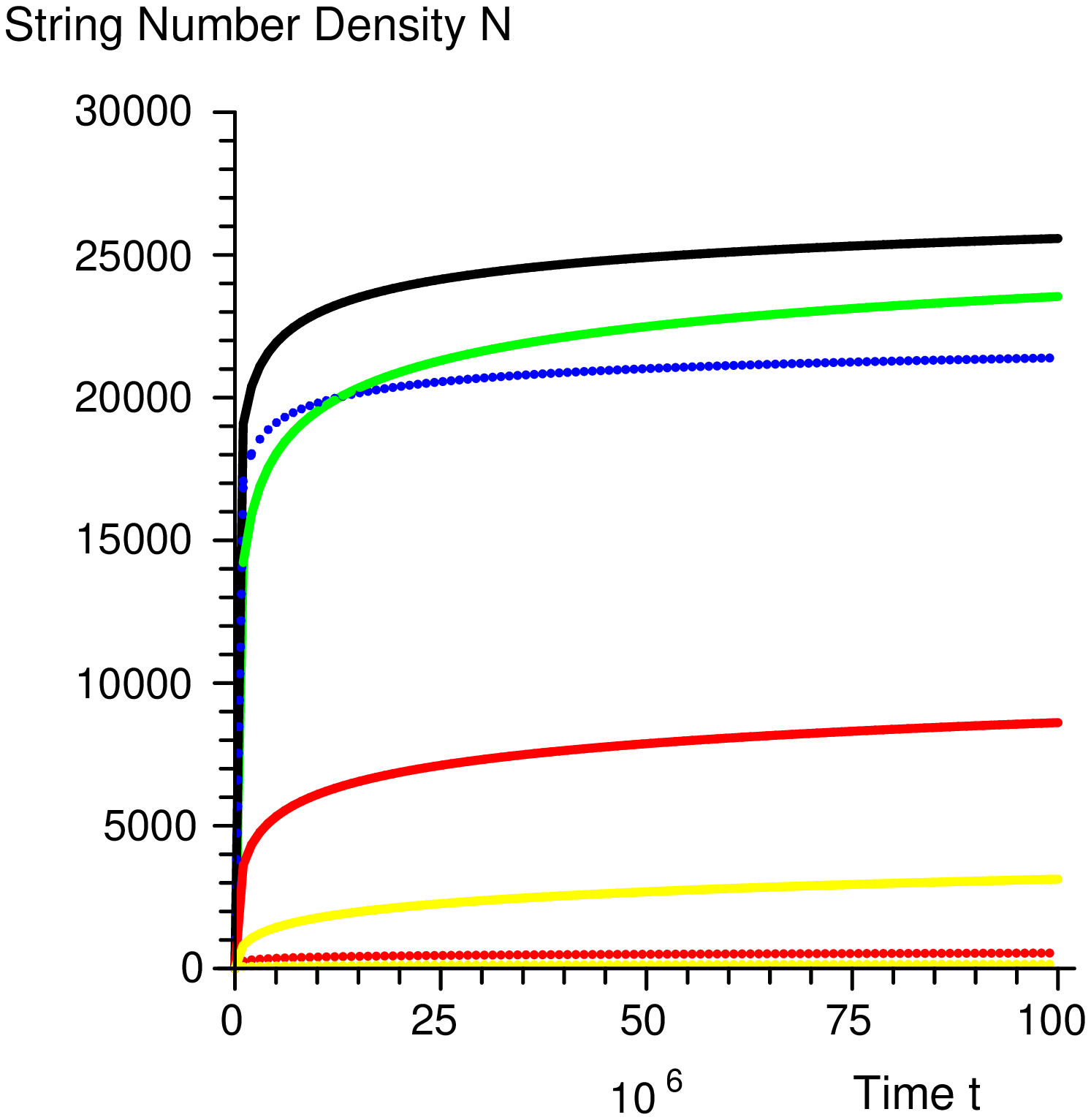}
    \caption{\label{super_rho} Evolution of normalised string density
             $N=\rho t^2/\mu$ for the lightest components of a cosmic
             superstring network.  F-strings are shown with a black
             solid line, D-strings with dotted blue, $(1,\pm 1)$ strings
             are in green, $(2,\pm 1)$ in red and $(3,\pm 1)$ in yellow.
             Type $(1,\pm 2)$ and $(1,\pm 3)$ are shown in dotted red and
             dotted yellow respectively.  All string components are
             seen to approach scaling with heavier bound states less
             abundant.  As one reduces the intercommuting probability
             (left to right) the string densities increase and light
             strings become more dominant.}
   \end{center} 
  \end{figure}

 Fig.~\ref{super_rho} shows the scaling density results of a model  
 containing bound string states up to $(3,\pm 1)$ and $(1,\pm 3)$.  
 In agreement with Ref.~\cite{MTVOS} the string densities of heavier 
 states fall rapidly as the string tension increases.  This is because 
 of equation (\ref{Paddsubtr}), which tends to give a small value of 
 $P^+$ (hence a large value of $P^-$) for large $p$'s and $q$'s, so 
 that heavy strings have the tendency to break to lighter ones.  It 
 is therefore possible to obtain a very good approximation of such a 
 network by truncating the equations at a relatively low $s\!=\!p+q$, 
 like, in this case, $s=4$.  As one reduces the intercommuting probability 
 (Fig.~\ref{super_rho}) string densities increase, as expected, and the 
 fall of density with increasing string tension becomes more prominent.  
 The relative importance of light to heavier string states depends on 
 both the string coupling $g_s$ (appearing in the string tensions and 
 $P^\pm$), and the intercommuting probability absorbed in $\tilde c$'s 
 and $\tilde d$'s.  However, the generic behaviour is that the string 
 number density is dominated by strings of type $(1,0)$, $(0,1)$ and 
 $(1,\pm 1)$, all having comparable number densities, while higher 
 composite states are heavily suppressed.  

 The fact that the number density of the first bound state $(1,\pm 1)$ 
 is comparable to that of the unbound strings can be understood in terms 
 of the difference in string tension between F- and D-strings.  This 
 makes it energetically favourable for the light $(1,0)$ strings to be 
 in a bound $(1,\pm 1)$ state.   One can then envisage such a network 
 as being predominantly composed by a heavy component of D-strings, 
 together with a `cobweb' of light F-strings, which tend to stick to 
 the D-string network, giving rise to another heavy FD-string component.   

\begin{figure}[h]
   \begin{center}
    \includegraphics[height=2.9in,width=3.2in]{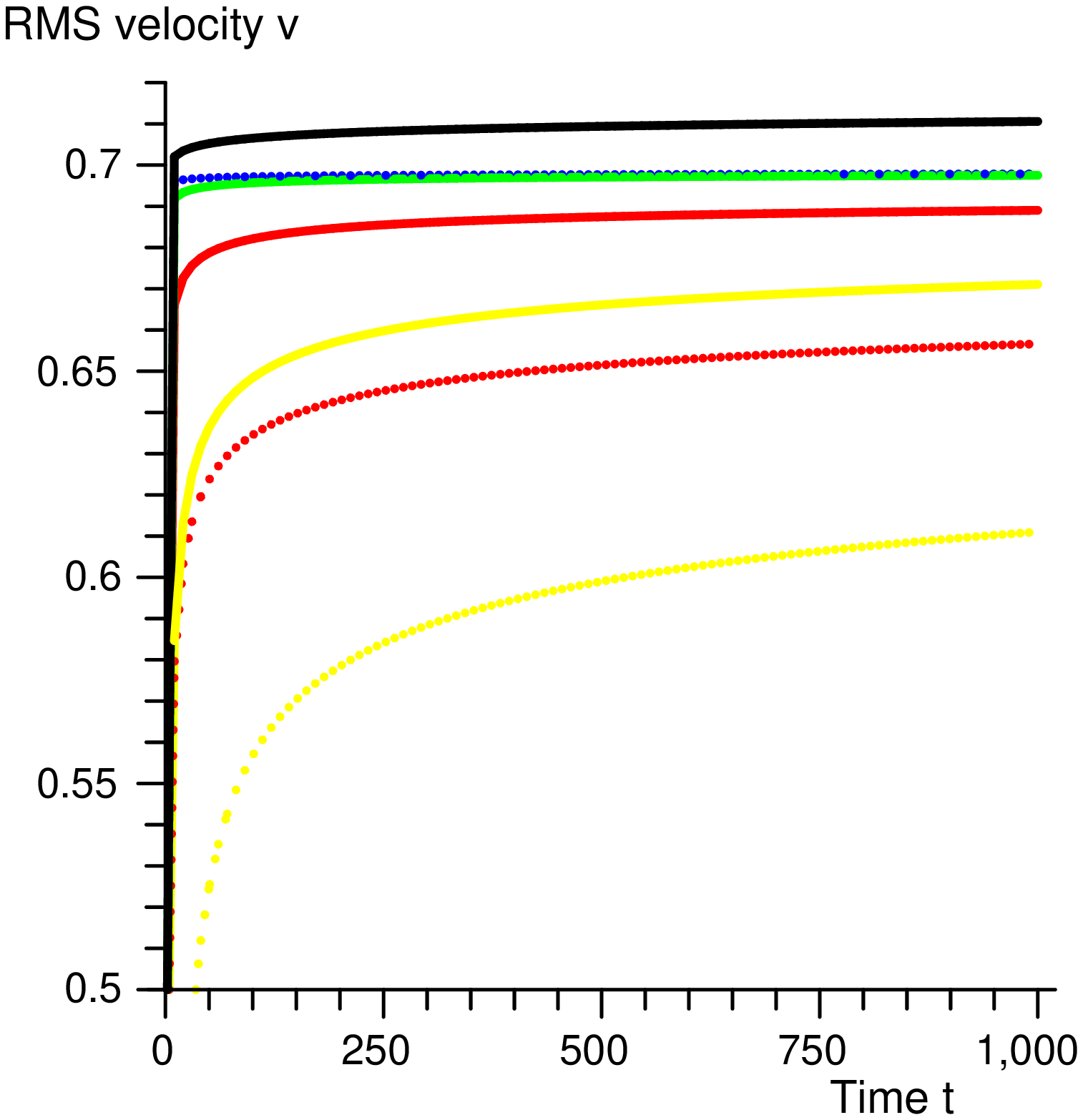}
    \includegraphics[height=2.9in,width=3.2in]{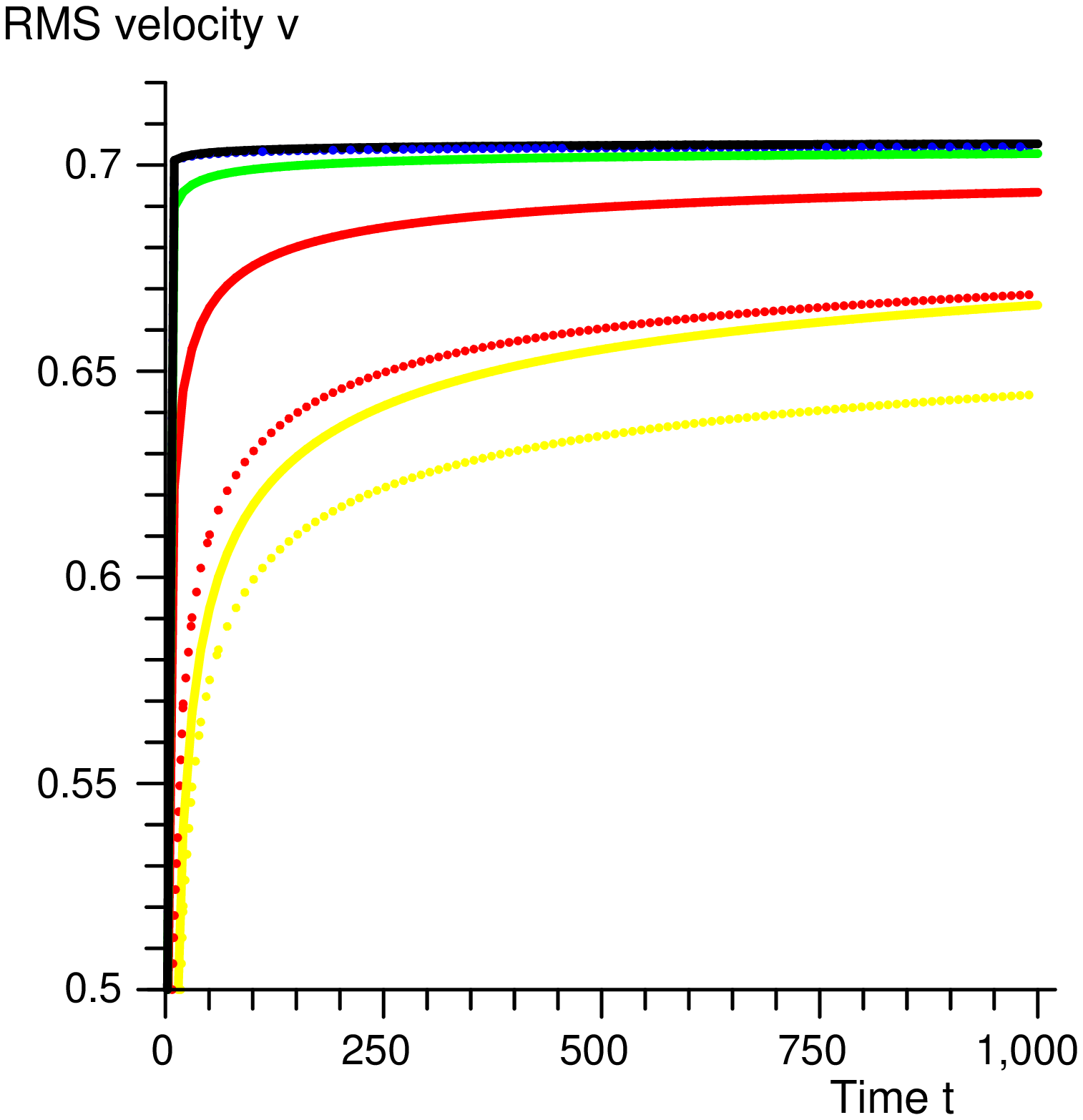}
    \caption{\label{super_v}  Evolution of the rms string velocity
             $v$ for the first lightest components of a cosmic
             superstring network.  F-strings are shown with a black    
             solid line, D-strings with dotted blue, $(1,\pm 1)$ strings
             are in green, $(2,\pm 1)$ in red and $(3,\pm 1)$ in yellow.
             The dotted red and yellow curves correspond to strings of 
             type $(1,\pm 2)$ and $(1,\pm 3)$.  All string components 
             asymptotically reach a constant value, which increases as 
             one reduces the intercommuting probability (left to right).}
   \end{center}
  \end{figure}

 In our velocity dependent model we separately evolve the rms velocities 
 of each string component of the network, which, since each string type 
 has a different correlation length, can be significantly different.   
 This is shown in Fig.~\ref{super_v} where, evidently, the less abundant  
 strings also have smaller velocities.  This could in principle be  
 observable as the velocity enters the temperature discontinuity  
 through the Kaiser-Stebbins effect \cite{KaisSteb,book}.         

 The above results provide further evidence in favour of the possibility  
 of scaling in cosmic superstring networks and other `entangled' 
 networks with junctions.  Particularly interesting is the fact 
 that the scaling solutions we found do not rely on any additional 
 energy damping mechanism, apart from loop production through 
 self-intersections of strings.  Indeed, by imposing energy conservation 
 during zipping events, we have ensured that no energy is damped 
 during the zipping processes, which would facilitate the reaching
 of a scaling regime.  Possible non-elasticity of the zipping events
 would provide an additional decay channel, which would therefore 
 further support our scaling results.  One should keep in mind  
 that the models we presented here and applied to the case of cosmic 
 superstrings, are phenomenological in nature and, as such, do not 
 model in detail a number of microphysical processes, related, for 
 example, to the dynamics of string junctions~\cite{BetLagMat}  
 (see Ref.~\cite{RajSakSto} for recent field theory simulations).  
 However, progress in understanding the complex, non-linear evolution 
 of cosmic strings has been possible with the combination of different 
 approaches (microphysical models, Nambu-Goto and field theory simulations, 
 phenomenological models), and there is already strong evidence 
 that models like those presented here capture most of the relevant 
 physics.  Being able to construct such analytic models of 
 multi-tension string evolution and come up with quantitative  
 predictions is an important step in our understanding of non-abelian  
 cosmic (super)string evolution.  It would be interesting to combine  
 these models with other approaches to cosmic superstring modelling   
 as for example field theory models \cite{CopSaf,Saffin,HindSaf}  
 or Nambu-Goto multi-tension simulations.

\section{\label{conc}Conclusion}   

 We have presented analytic velocity-dependent models for the cosmological  
 evolution of non-abelian string networks.  Apart from ordinary abelian 
 intercommutations, these models account for interactions between different 
 string types, producing Y-type junctions with linking segments stretching 
 between the originally colliding strings.  We have described such 
 interactions in terms of two limiting cases, namely zipper-type 
 interactions, where the link is produced by the zipping of the colliding 
 strings into a bound state, and bridge-type ones, where the link is a 
 newly formed string of a third type and the parent strings do not lose 
 significant string length.  A general Y-junction forming interaction 
 can be describing by a combination of these two inter-related mechanisms,
 and the relative weighting depends on the type of strings under study.  
 In non-abelian field theory, where the interactions are topologically 
 constrained, one can argue that the bridge-type picture is a good
 approximation as the colliding strings do not gain energy by converting
 length to the new type.  On the other hand the zipper interaction 
 corresponds closely to the case of cosmic superstrings, where D-strings 
 and F-strings tend to bind together to form heavier composites.    

 In the bridge case, we have applied our models to the evolution of $Z_N$ 
 string networks and found that the presence of Y-type junctions does  
 not generally lead to string frustration.  Instead, scaling solutions  
 exist for a wide range of physically relevant choices of parameters.   
 Here, we have thus discussed the physical properties and scaling 
 behaviour of $Z_N$ networks for $N=3,4,...,7$ noting the same 
 qualitative behaviour.       
 
 In the zipper case, we have modelled a cosmic superstring network as 
 one consisting of two types of elementary strings (corresponding to 
 F- and D-strings) which can zip together to form bound states between  
 $p$ strings of one type and $q$ strings of the other.  We have  
 demonstrated scaling of all string types, with heavier strings 
 generally less populated than lighter ones, as noted in Ref.~\cite{MTVOS}.
 We have also obtained the scaling velocities of each network component, 
 with heavier strings moving slower than lighter ones.  The general 
 picture emerging from this analysis, is a string network whose number 
 density is dominated by the lightest (unbound) strings and the first 
 bound state between them, with all heavier bound states being suppressed.  
 The first bound state develops a comparable number density to the unbound 
 strings, even though it has a higher string tension.  This is to be 
 understood in terms of the difference in string tension between F- and 
 D-strings, which makes it energetically favourable for the lightest 
 F-strings to be in a bound FD-state.  Thus, a scaling superstring network 
 is dominated by D-strings and the first FD bound state, with a cobweb 
 of lighter F-strings of comparable number density, but subdominant 
 energy density.        

 The strength of our scaling results stems from the fact that we have 
 not allowed additional energy damping processes associated to the 
 production of junctions (though our models can accomodate these 
 possibilities also), as we have enforced energy conservation 
 in the relevant string interaction events.  Thus, first, we can be 
 sure that we are not observing a `spurious' scaling behaviour due 
 to artificially throwing away some of the energy involved, and, 
 second, our scaling results provide evidence that loop production 
 alone may be sufficient for scaling (even in the presence of 
 junctions) without the necessity of extra energy losses, as 
 for example massive radiation at zipping.  However, it should be 
 highlighted that our findings provide evidence, rather than proof, 
 of scaling in such networks.  Indeed, our models are phenomenological 
 in nature and, as such, have their own limitations.  In particular, 
 the detailed microphysics of junctions, for example the possibility
 of unzipping, is not modelled explicitly, but only statistically 
 and indirectly, through choices of the relevant parameters.  In 
 principle, if juncion dynamics is sufficiently complicated, these
 coefficients could inherit time dependence, a possibility we have 
 not considered.  One should therefore combine our results with 
 other complementary approaches to this difficult problem, each 
 of which captures only part of the relevant physics.  What is 
 encouraging, is that all complementary approaches seem to independently 
 point towards scaling in these networks.    

 It would be interesting to compare these models to field theory simulations 
 \cite{CopSaf,Saffin,HindSaf} and see whether quantitative agreement can  
 be established.  It would also be desirable to generalise them to  
 include a second length-scale for each network component, as, for  
 small intercommuting probabilities likely to appear in cosmic  
 superstring networks, a two-scale VOS model (see also  
 Ref.~\cite{AusCopKib} for a three-scale model) is needed in 
 order to match Nambu-Goto simulations of single string networks  
 \cite{intprob}.  Other variants of our models could also be constructed.  
 For example, the models presented here assume that the production of 
 junctions can be approximated by the two mechanisms described in 
 Ref.~\cite{McGraw1}, namely the `zipper', where the string lengths 
 lost from each of the zipping strings equal the length of the produced 
 link, and the `bridge', where the lengths of the colliding strings are 
 preserved.  Recent studies of junction formation based on the Nambu-Goto 
 action \cite{CopKibSteer1,CopKibSteer2} suggest a slightly different 
 mechanism where strings lose or gain length subject to a constraint, 
 which the vertex has to satisfy.  This mechanism could be readily 
 accommodated in our models.  It is an interesting question to what  
 extent scaling results depend on such choices.  Analytic models like 
 these will be useful in comparing and contrasting with other  
 approaches to `entangled' network evolution.  Although quantitative  
 agreement with numerical simulations has not yet been investigated, 
 these models provide further evidence that scaling behaviour in 
 networks with junctions is possible, if not generic.

\begin{acknowledgments}
We would like to thank Carlos Martins for many fruitful
conversations and valuable suggestions.  We also thank Alkistis 
Pourtsidou for pointing out a typographical mistake on the Hubble 
expansion term of the string density equations, which was corrected 
in the current version.  This work has also
benefited from discussions with Ed Copeland and Anne Davis.  A.A.
acknowledges financial support from the Cambridge European Trust,
the Cambridge Newton Trust and the European Network on Random
Geometry (ENRAGE).  This work was also supported by PPARC.
\end{acknowledgments}

\newpage 
\appendix
 \section{\label{appendix}Parameter matrices for $Z_N$ strings,
                          $4 \le N \le 7$}

  In this appendix we explore the topologically allowed outcomes from
  string collisions in the case of $Z_4$, $Z_5$, $Z_6$ and $Z_7$ networks.

  \subsection{\label{app_Z4}$Z_4$ strings}
   Consider $Z_4$ with generator $h$ and elements $h_i=1,h,h^2,h^3$.
   There are three types of string (say type 1, 2 and 3) corresponding
   to the non-trivial group elements $h$, $h^2$ and $h^3$ respectively.
   We have $h\cdot h=h^2$ so that strings of type 1 can produce a type
   2 segment when they interact, that is
   \be
    1+1\rightarrow 2 .
   \ee
   Thus, the crossing of two type 1 strings can have two different
   outcomes, as shown:
   \begin{figure}[h]
    \includegraphics[height=2.5cm,width=7.5cm]{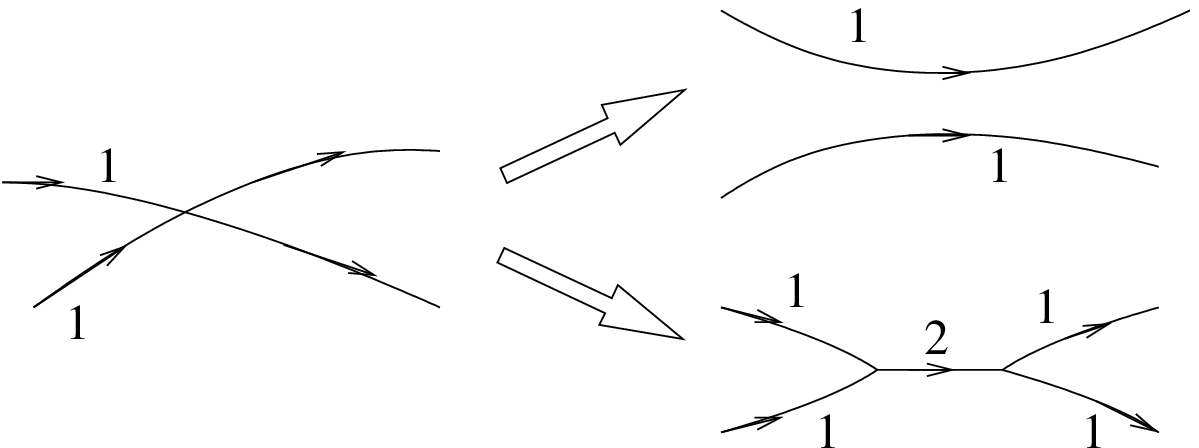}
   \end{figure}

   Similarly, we have:
   $h\cdot h^2=h^3 \Rightarrow 1+2 \rightarrow 3,$
   corresponding to
   \begin{figure}[h]
    \includegraphics[height=1.5cm,width=7.5cm]{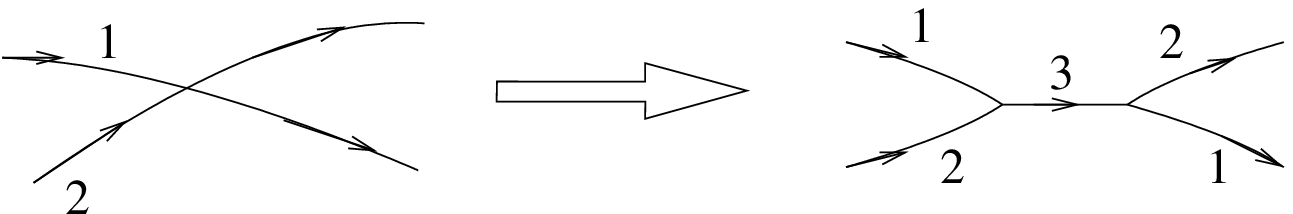}
   \end{figure}

   $h^2\cdot h^2=1 \Rightarrow 2+2 \rightarrow \emptyset,$
   meaning that two incoming type 2 strings can end on a vertex and
   so the interaction between type 2 strings can then have two
   outcomes:
   \begin{figure}[h]
    \includegraphics[height=2.5cm,width=7.5cm]{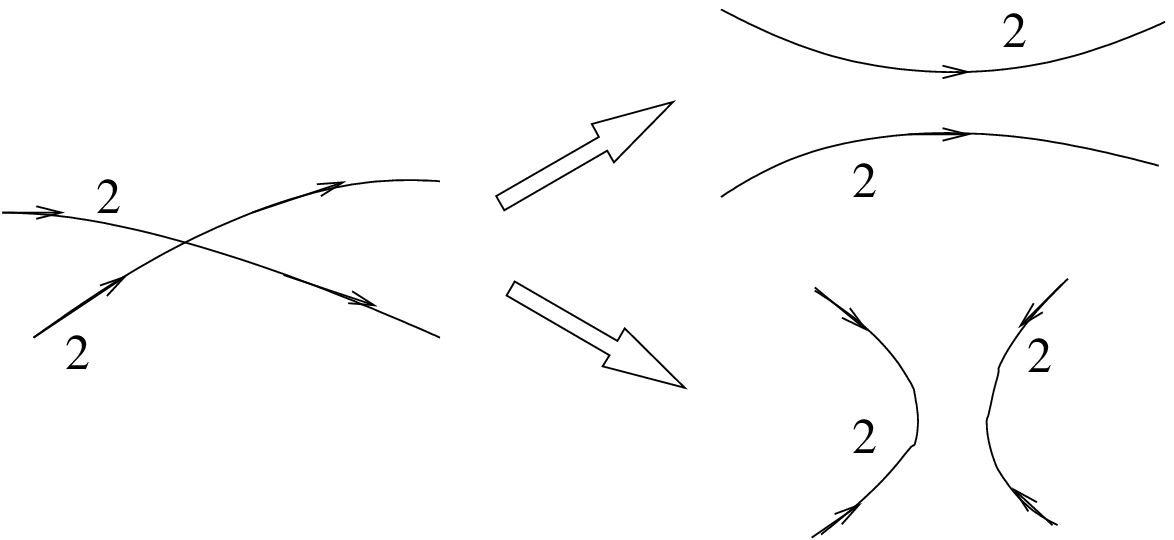}
   \end{figure}

   $h^3\cdot h^3=h^2 \Rightarrow 3+3 \rightarrow 2,$
   which is as in the first diagram, \\
   and:
\newpage
   $h\cdot h^3=1 \Rightarrow 1+3 \rightarrow \emptyset,$
   that is a type 1 and and a type 3 strings can end on a vertex,
   and can be treated as being the same string type with opposite
   orientation:
   \begin{figure}[h]
    \includegraphics[height=1.5cm,width=6cm]{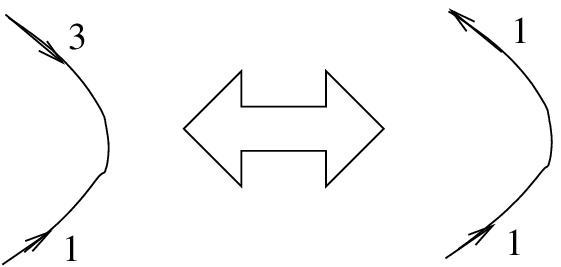}
   \end{figure}

   Hence there are only two distinct types of string, type $1\!\equiv\!-3$
   and $2$, with type $2$ being self-conjugate.  The parameter matrix
   corresponding to the above interactions is (see also section
   \ref{mod_ZN}):
   \be\label{matrix_Z4}
    {\cal M}=
    \left( \begin{array}{cc} (\tilde c_1, \tilde d^2_{11}) &
                             (\tilde d_{12}^1, 0)             \\
                             (\tilde d_{21}^1, 0)          &
                             (0, \tilde c_2)
           \end{array}
    \right)  .
   \ee

  \subsection{\label{app_Z5}$Z_5$ strings}
   For $Z_5$ we have $h_i=1,h,h^2,h^3,h^4$.  Associating a string type
   to each non-trivial generator, as before, we have:
   \be\nonumber
    \left . \begin{array}{cc}  1+1 \rightarrow 2 & \\
                               1+2 \rightarrow 3 & \\
                               1+3 \rightarrow 4 & \\
                               1+4 \rightarrow \emptyset &
                                    \Rightarrow 1 \equiv -4 \\
                               2+2 \rightarrow 4 & \\
                               2+3 \rightarrow \emptyset &
                                    \Rightarrow 2 \equiv -3 \\
                               2+4 \rightarrow 1 & \\
                               3+3 \rightarrow 1 & \\
                               3+4 \rightarrow 2 & \\
                               4+4 \rightarrow 3 & \\
            \end{array} \right\}
            \Rightarrow \begin{array}{ccc}
                          1+1 & \rightarrow &  2  \\
                          1+2 & \rightarrow & -2  \\
                          1 + (-2) & \rightarrow & -1  \\
                          2+2 & \rightarrow & -1  \\
                        \end{array}
   \ee
   Again, there are two types of string with interaction diagrams: \\
   $1+1 \rightarrow 2$
   \begin{figure}[h]
    \includegraphics[height=2.5cm,width=7.5cm]{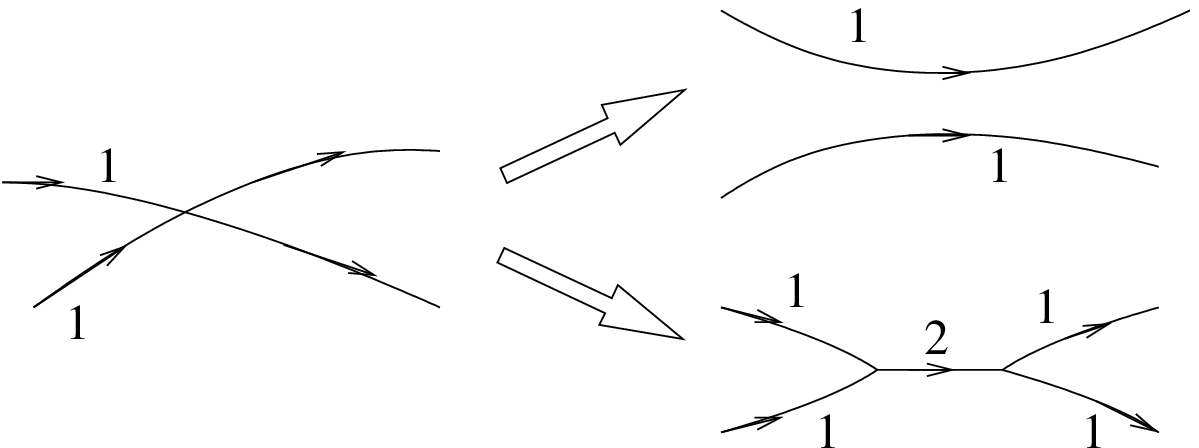}
   \end{figure}
\newpage
   $1+2 \rightarrow -2$
   \begin{figure}[h]
    \includegraphics[height=1.5cm,width=6.5cm]{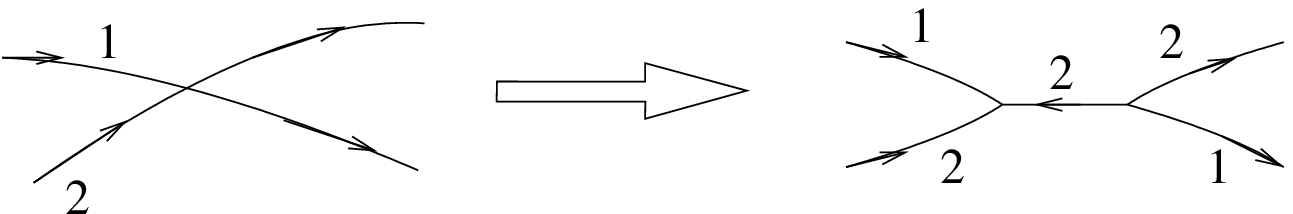}
   \end{figure}

   $1+(-2) \rightarrow -1$
   \begin{figure}[h]
    \includegraphics[height=1.5cm,width=7.5cm]{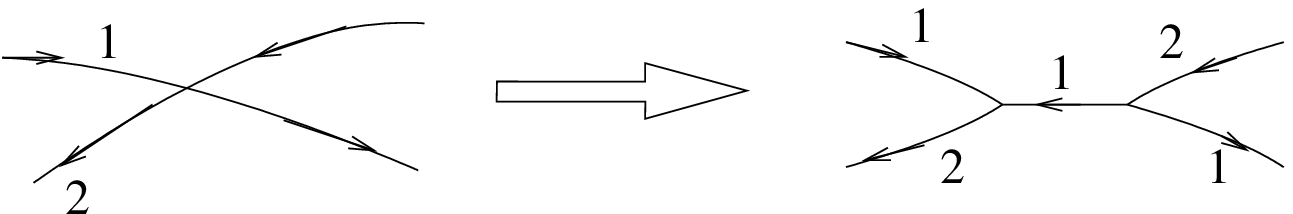}
   \end{figure}

   $2+2 \rightarrow -1$
   \begin{figure}[h]
    \includegraphics[height=2.5cm,width=7.5cm]{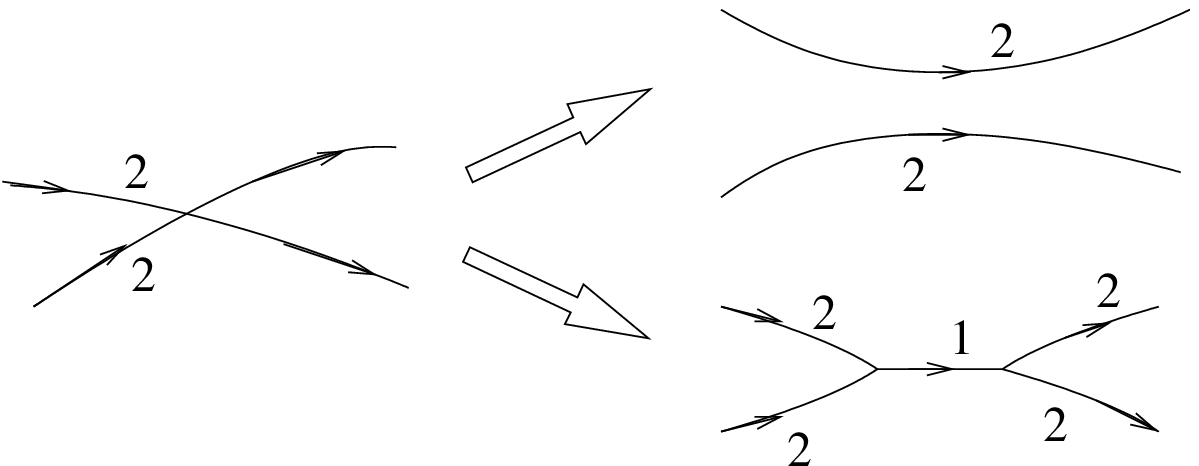}
   \end{figure} \\

   These are described by the parameter matrix:
   \be\label{matrix_Z5}
    {\cal M}=
    \left( \begin{array}{cc} (\tilde c_1, \tilde d^2_{11})      &
                             (\tilde d_{12}^1, \tilde d_{12}^2)    \\
                             (\tilde d_{21}^1, \tilde d_{21}^2) &
                             (\tilde d_{22}^1, \tilde c_2)
           \end{array}
    \right)  .
   \ee
   Note that there is no self-conjugate string in this case.

   \subsection{\label{app_Z6}$Z_6$ strings}
   For $Z_6$ we have $h_i=1,h,h^2,h^3,h^4,h^5$. Now:
   \be\nonumber
    \left . \begin{array}{cl}  1+1 \rightarrow 2 & \\
                               1+2 \rightarrow 3 & \\
                               1+3 \rightarrow 4 & \\
                               1+4 \rightarrow 5 & \\
                               1+5 \rightarrow \emptyset &
                                    \Rightarrow 1 \equiv -5 \\
                               2+2 \rightarrow 4 & \\
                               2+3 \rightarrow 5 & \\
                               2+4 \rightarrow \emptyset &
                                    \Rightarrow 2 \equiv -4 \\
                               2+5 \rightarrow 1 & \\
                               3+3 \rightarrow \emptyset &
                                    \Rightarrow 3 \,\,
                                    \mathrm{self}\!-\!
                                         \mathrm{conjugate} \\
                               3+4 \rightarrow 1 & \\
                               3+5 \rightarrow 2 & \\
                               4+4 \rightarrow 2 & \\
                               4+5 \rightarrow 3 & \\
                               5+5 \rightarrow 4 & \\
            \end{array} \right\}
            \Rightarrow \begin{array}{ccc}
                          1+1 & \rightarrow &  2  \\
                          1+2 & \rightarrow &  3  \\
                          1+3 & \rightarrow & -2  \\
                          1 + (-2) & \rightarrow & -1  \\
                          2+2 & \rightarrow & -2  \\
                          2+3 & \rightarrow & -1  \\
                          2 + (-1) & \rightarrow & -1  \\
                          3 + (-2) & \rightarrow & 1  \\
                          3 + (-1) & \rightarrow & 2  \\
                        \end{array}
   \ee
   Some of the new diagrams arising are:   \\
   $1+2 \rightarrow 3$     \\
   $1+(-2) \rightarrow -1$
   \begin{figure}[h]
    \includegraphics[height=2.7cm,width=7.5cm]{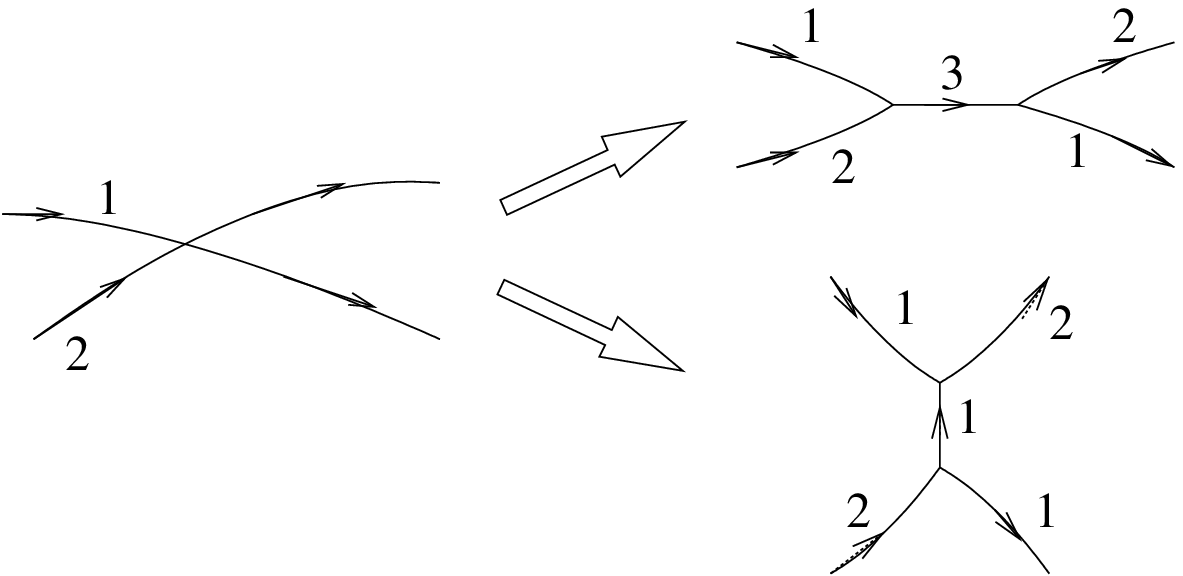}
   \end{figure} \\
   and:   \\
   $2+2 \rightarrow -2$
   \begin{figure}[h]
    \includegraphics[height=2.2cm,width=7.5cm]{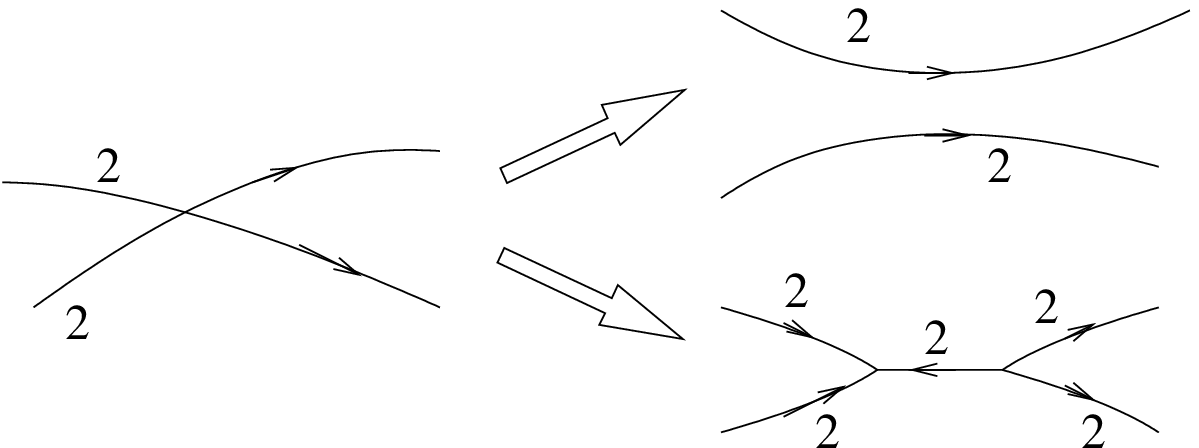}
   \end{figure}
\newpage
   There are only three string types with parameter matrix:
   \be\label{matrix_Z6}
    {\cal M}=
    \left( \begin{array}{ccc} (\tilde c_1, \tilde d^2_{11},0)       &
                             (\tilde d_{12}^1,0,\tilde d_{12}^3)    &
                             (0,\tilde d_{13}^2,0)                   \\
                             (\tilde d_{12}^1,0,\tilde d_{12}^3)    &
                             (0,{}_{\tilde d_{22}^2}^{\tilde c_2},0)&
                             (\tilde d_{23}^1,0,0)                   \\
                             (0,\tilde d_{13}^2,0)                  &
                             (\tilde d_{23}^1,0,0)                  &
                             (0, 0, \tilde c_3)
           \end{array}
    \right)  .
   \ee
   String $3$ is self-conjugate.

   \subsection{\label{app_Z7}$Z_7$ strings}
   For $Z_7$ we have $h_i=1,h,h^2,h^3,h^4,h^5,h^6$. Working as before:
   \be\nonumber
    \left . \begin{array}{cl}  1+1 \rightarrow 2 & \\
                               1+2 \rightarrow 3 & \\
                               1+3 \rightarrow 4 & \\
                               1+4 \rightarrow 5 & \\
                               1+5 \rightarrow 6 & \\
                               1+6 \rightarrow \emptyset &
                                    \Rightarrow 1 \equiv -6 \\
                               2+2 \rightarrow 4 & \\
                               2+3 \rightarrow 5 & \\
                               2+4 \rightarrow 6 & \\
                               2+5 \rightarrow \emptyset &
                                    \Rightarrow 2 \equiv -5 \\
                               2+6 \rightarrow 1 & \\
                               3+3 \rightarrow 6 & \\
                               3+4 \rightarrow \emptyset &
                                    \Rightarrow 3 \equiv -4 \\
                               3+5 \rightarrow 1 & \\
                               3+6 \rightarrow 2 & \\
                               4+4 \rightarrow 1 & \\
                               4+5 \rightarrow 2 & \\
                               4+6 \rightarrow 3 & \\
                               5+5 \rightarrow 3 & \\
                               5+6 \rightarrow 4 & \\
            \end{array} \right\}
            \Rightarrow \begin{array}{ccc}
                          1+1 & \rightarrow &  2  \\
                          1+2 & \rightarrow &  3  \\
                          1 + (-2) & \rightarrow &  -1  \\
                          1+3 & \rightarrow & -3  \\
                          1 + (-3) & \rightarrow &  -2  \\
                          2+2 & \rightarrow & -3  \\
                          2+3 & \rightarrow & -2  \\
                          2 + (-3) & \rightarrow & -1  \\
                          3+3 & \rightarrow & -1  \\
                          3 + (-2) & \rightarrow & 1  \\
                          3 + (-1) & \rightarrow & 2  \\
                        \end{array}
   \ee
   Again, there are only three string types, but no self-conjugate strings.
   The parameter matrix is now:
   \be\label{matrix_Z7}
    {\cal M}=
    \left( \begin{array}{ccc} (\tilde c_1, \tilde d^2_{11},0)      &
                             (\tilde d_{12}^1,0,\tilde d_{12}^3)   &
                             (0,\tilde d_{13}^2,\tilde d_{13}^3)    \\
                             (\tilde d_{12}^1,0,\tilde d_{12}^3)   &
                             (0, \tilde c_2, \tilde d_{22}^3 )&
                             (\tilde d_{23}^1,\tilde d_{23}^2,0)    \\
                             (0,\tilde d_{13}^2,\tilde d_{13}^3)   &
                             (\tilde d_{23}^1,\tilde d_{23}^2,0)   &
                             (\tilde d_{33}^1, 0, \tilde c_3)
           \end{array}
    \right)  .
   \ee

\bibliography{NAVOS}

\end{document}